%% file: main-arxiv.tex
\documentclass[11pt,letter]{article}

\usepackage{graphicx}
\usepackage{multirow}
\usepackage{lineno}
\usepackage[normalem]{ulem}
\usepackage{color}
\usepackage[colorlinks = true,
            linkcolor = blue,
            urlcolor  = blue,
            citecolor = blue,
            anchorcolor = blue]{hyperref}

\usepackage{xspace}
\usepackage{rotating,booktabs}
\usepackage{amsmath,amssymb}
\usepackage{cite}

\input{workshopsymbols}

\newcommand{\chapter}[1]{%
\title{\parbox{0.175\linewidth}{\includegraphics[width=\linewidth]{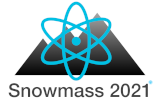}} \hfill \parbox{0.3\linewidth}{\normalsize \hfill September 16, 2022} \\ \vspace{0.06\textheight} \textbf{\huge #1} \vspace{0.02\textheight} \\ 
\textbf{\small Report of Topical Group RF1} \\
\textbf{\small Rare Processes and Precision Measurements Frontier} \vspace{0.02\textheight}
}
}

\newcommand{\authorlist}[2]{%
\author{#1 \\[0.2in] \small{#2}}
\date{}
\maketitle
}

\newenvironment{execsummary}{\begin{abstract}}{\end{abstract}\newpage\tableofcontents\cleardoublepage}

\begin{document}

\input{Rare/RF01/bcquarks}

\end{document}

%% file: workshopsymbols.tex
\def\ie{{\it i.e.}}
\def\eg{{\it e.g.}}
\def\etc{{\it etc.}}
\def\etal{{\it et al.}}

\def\beq{\begin{equation}}
\def\eeq#1{\label{#1}\end{equation}}
\def\eeqn{\end{equation}}

\newenvironment{Eqnarray}%
   {\arraycolsep 0.14em\begin{eqnarray}}{\end{eqnarray}}
\def\beqa{\begin{Eqnarray}}
\def\eeqa#1{\label{#1}\end{Eqnarray}}
\def\eeqan{\end{Eqnarray}}

\let\bar=\overbar

\def\lsim{\mathrel{\raise.3ex\hbox{$<$\kern-.75em\lower1ex\hbox{$\sim$}}}}
\def\gsim{\mathrel{\raise.3ex\hbox{$>$\kern-.75em\lower1ex\hbox{$\sim$}}}}

\def\del{\partial}
\def\Dslash{\not{\hbox{\kern-4pt $D$}}}
\def\dslash{\not{\hbox{\kern-2pt $\del$}}}
\def\pslash{\not{\hbox{\kern-2pt $p$}}}
\def\ETmiss{\not{\hbox{\kern-4pt $E$}}_T}

\def\Dlr{\mathrel{\raise1.5ex\hbox{$\leftrightarrow$\kern-1em\lower1.5ex\hbox{$D$}}}}

\def\MSB{{\bar{M \kern -2pt S}}}
\def\msb{{\bar{\scriptsize M \kern -1pt S}}}

\def\drb{{\bar{\scriptsize D \kern -1pt R}}}



%% file: Rare/RF01/bcquarks.tex


\input{Rare/RF01/macros}

\chapter{\boldmath Weak Decays of $b$ and $c$ Quarks}

\authorlist{Angelo Di Canto$^1$, Stefan Meinel$^2$}{$^1$\textit{Brookhaven National Laboratory}, $^2$\textit{University of Arizona}}

\begin{execsummary}
Precision measurements in weak decays of heavy flavored hadrons can test in unique ways our understanding of the fundamental interactions and of the observed baryon asymmetry in the Universe. The high sensitivity of such decays to beyond-Standard-Model physics, combined with the lack of major discoveries in direct production of new particles, motivates the continuation of a strong heavy-flavor program in the next decades. The observation of several anomalies by the BaBar, Belle and LHCb experiments in such decays, including evidence for violation of lepton universality, provides particular motivation to vigorously pursue this program. While the mass scales probed by direct searches for non-Standard-Model phenomena at the energy frontier will only marginally increase in the near future, a substantial advancement is expected in the study of weak decays of $b$ and $c$ quarks. The next 10 to 20 years will see the development of a highly synergistic program of experiments at both $pp$ and $e^+e^-$ colliders. This program will be complemented by advancements in theory, including both lattice and continuum calculations. Experimental measurements and theory predictions of several key observables will reach unprecedented precision and will allow to test the Standard Model in ways that have not been possible thus far. With a strong participation in this program, the US high-energy-physics community will remain at the forefront of indirect searches for new physics and retain its leading role in expanding humankind's understanding of fundamental interactions.
\end{execsummary}

\section*{Introduction\addcontentsline{toc}{section}{Introduction}}
This report describes the physics case for precision studies of weak decays of $b$ and $c$ quarks, and it discusses the experimental and theory programs needed to exploit these physics opportunities in the next decades. It is based on the several white papers submitted by the community and, in particular, on Refs.~\cite{LFUV-Overview-Whitepaper,Rare-decay-Overview-Whitepaper,CKM-Overview-Whitepaper,CPV-Overview-Whitepaper} -- which provide an overview of the contributions discussing weak decays of $b$ and $c$ quarks. This report is not a review of heavy-quark physics, and no attempt has been made to provide complete references to prior work. Section~\ref{sec:discovery} discusses the motivation for heavy-quark physics, giving an overview of its unique potential for discoveries of new dynamics up to very high energy scales. Section~\ref{sec:experiments} presents the experimental efforts planned/proposed over the 2020-2030 decades, followed in Sec.~\ref{sec:far-future} by a discussion of the opportunities for a continued heavy-quark-physics program at facilities further into the future. The expected theory progress is then outlined in Sec.~\ref{sec:theory-progress}. The reports concludes in Sec.~\ref{sec:USinvolvement} with our recommendation to ensure a strong involvement of the U.S. high-energy-physics (HEP) community in this field of research.

\section{The path to discovery in heavy-quark physics\label{sec:discovery}}
The power of indirect searches for new fundamental physics in rare processes and precision measurements is rooted in the basic principles of quantum field theory. The probability amplitude for the transition from a certain initial state to a certain final state is the sum of all possible Feynman diagrams with these initial and final states. Crucially, the internal (virtual) particles in Feynman diagrams are not required to be on the mass shell, which means that arbitrarily heavy particles will contribute to the amplitude as long as there is no symmetry forbidding a coupling to them. The effects of the heavy $W^-$ bosons, for example, are seen in the $\beta$ decay of a neutron at much lower energy. Similarly, the charm, bottom, and top quarks were all predicted theoretically through their contributions as virtual particles to explain the observed phenomena of flavor-changing processes at lower energy, well before these particles could be produced directly. 

The study of weak decays of $b$ and $c$ quarks -- or more generally of quark-flavor physics -- has been essential in constructing the Standard Model (SM), and may very well also point us to what lies beyond. Many questions left unanswered by the SM, such as those about the origin of the large matter-antimatter asymmetry in the Universe, the mechanism giving neutrinos their masses, and the observed patterns and hierarchies in the many ``fundamental'' parameters, are directly related to flavor physics. More generally, proposed extensions of the SM, for example supersymmetry (which can alleviate the electroweak hierarchy problem and provides a dark-matter candidate), typically introduce new sources of flavor-changing interactions and new sources of charge-parity symmetry (\CP) violation. Flavor-physics measurements may reveal such effects, and can tightly constrain the parameter spaces of new theories (see, \eg, Refs.~\cite{Mahmoudi:2007gd,Akeroyd:2011kd,Atkinson:2022pcn}). Because the dependence on new particle masses and (flavor-violating) couplings is different than in the on-shell production, the new physics (NP) searches performed in weak decays of $b$ and $c$ quarks are also complementary to the direct searches at the energy frontier.

\begin{figure}[t]
\centering
\includegraphics[width=0.5\textwidth]{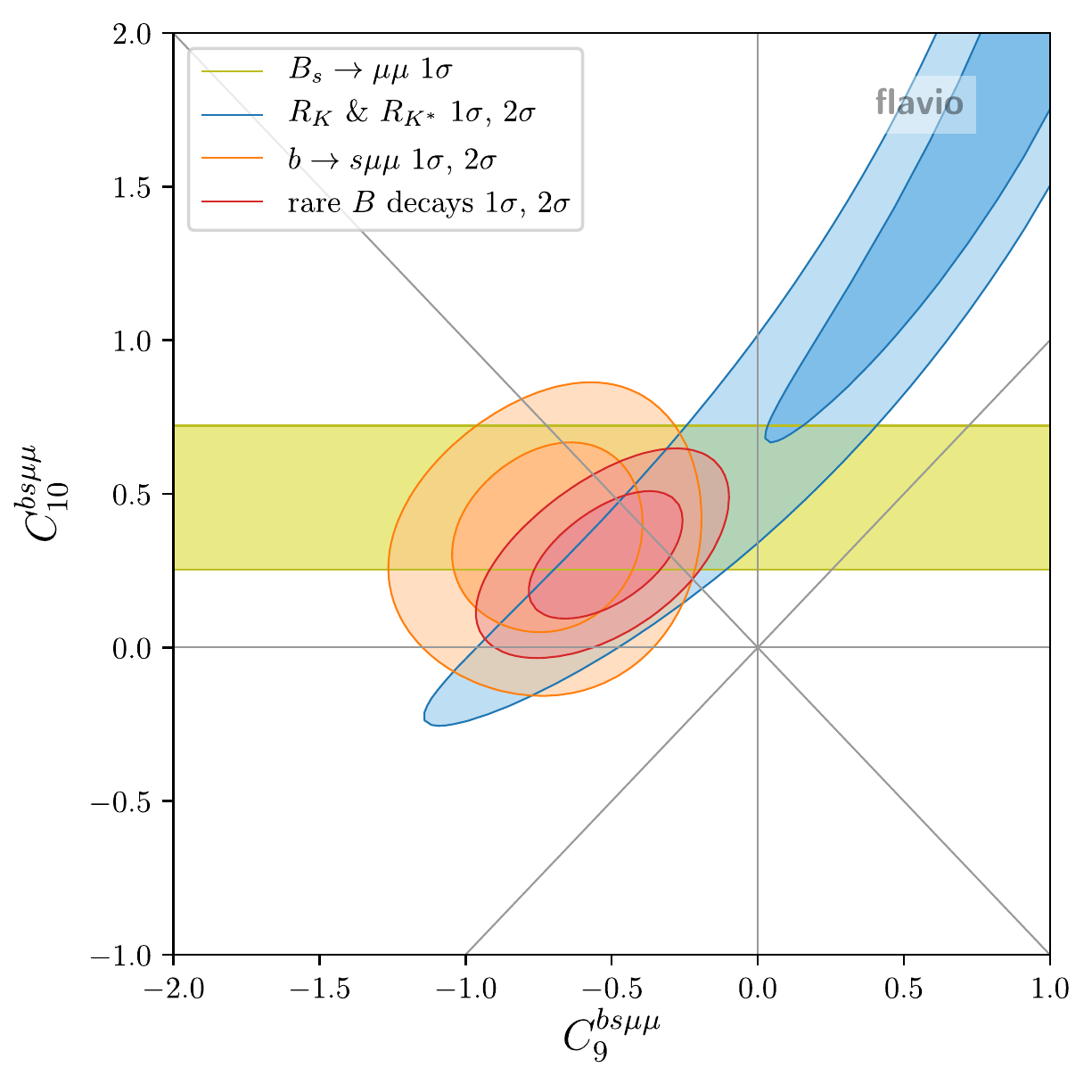}
\caption{Constraints at $1\sigma$ (darker) and $2\sigma$ (lighter) in the plane $C_9^{bs\mu\mu}$ \vs $C_{10}^{bs\mu\mu}$ resulting from $\BF(\Bs\to\mu^+\mu^-)$ (yellow-green), combination of the lepton-flavor-universality ratios $R_K$ and $R_{K^{*0}}$ (blue), combination of $b\to s\mu^+\mu^-$ observables (orange), and global fit of rare $b$ decays (red) \cite{Altmannshofer:2021qrr}. The Wilson coefficients $C_9^{bs\mu\mu}$ and $C_{10}^{bs\mu\mu}$ are the NP contributions to the couplings of the operators $O_9=(\bar{s}\gamma_\mu b_L)(\bar{\mu}\gamma^\mu\mu)$ and $O_{10}=(\bar{s}\gamma_\mu b_L)(\bar{\mu}\gamma^\mu\gamma_5\mu)$, respectively. The global fit result is inconsistent with the SM point (the origin) by $\sim5\sigma$.}\label{fig:C9C10}
\end{figure}

The unique properties and the richness of possible final states of weak decays of $b$ and $c$ quarks result in a particularly high potential for discovering new fundamental physics. As an example, rare $b$ and $c$ decays -- being strongly suppressed in the SM -- are potentially sensitive to very high NP scales of several 10's to 100\tev \cite{Rare-decay-Overview-Whitepaper}. Intriguingly, current experimental results for $b\to s \mu^+\mu^-$ branching fractions and angular observables, as well as ratios of $b\to s \mu^+\mu^-$ and $b\to s e^+e^-$ branching fractions, already show a coherent pattern of deviations from SM predictions~\cite{LFUV-Overview-Whitepaper}. The muon to electron ratios are predicted to be close to 1 in the SM with essentially no hadronic uncertainties, but are observed to be closer to 0.75 on average, suggesting violation of lepton-flavor universality. According to some analyses~\cite{Geng:2021nhg,Altmannshofer:2021qrr,Hurth:2021nsi,Alguero:2021anc}, fits to the experimental data and theory inputs yield pulls of $\gtrsim 5\sigma$ with respect to the SM, as shown for example in Fig.~\ref{fig:C9C10}. Moreover, restricting the fits to only the ratios of $b\to s \mu^+\mu^-$ and $b\to s e^+ e^-$ branching fractions along with $\BF(\Bs\to\mu^+\mu^-)$ still yields a significance for NP of $\gtrsim 4\sigma$~\cite{Geng:2021nhg,Altmannshofer:2021qrr,Hurth:2021nsi,Alguero:2021anc}.

\begin{figure}
\centering
\includegraphics[width=0.6\textwidth]{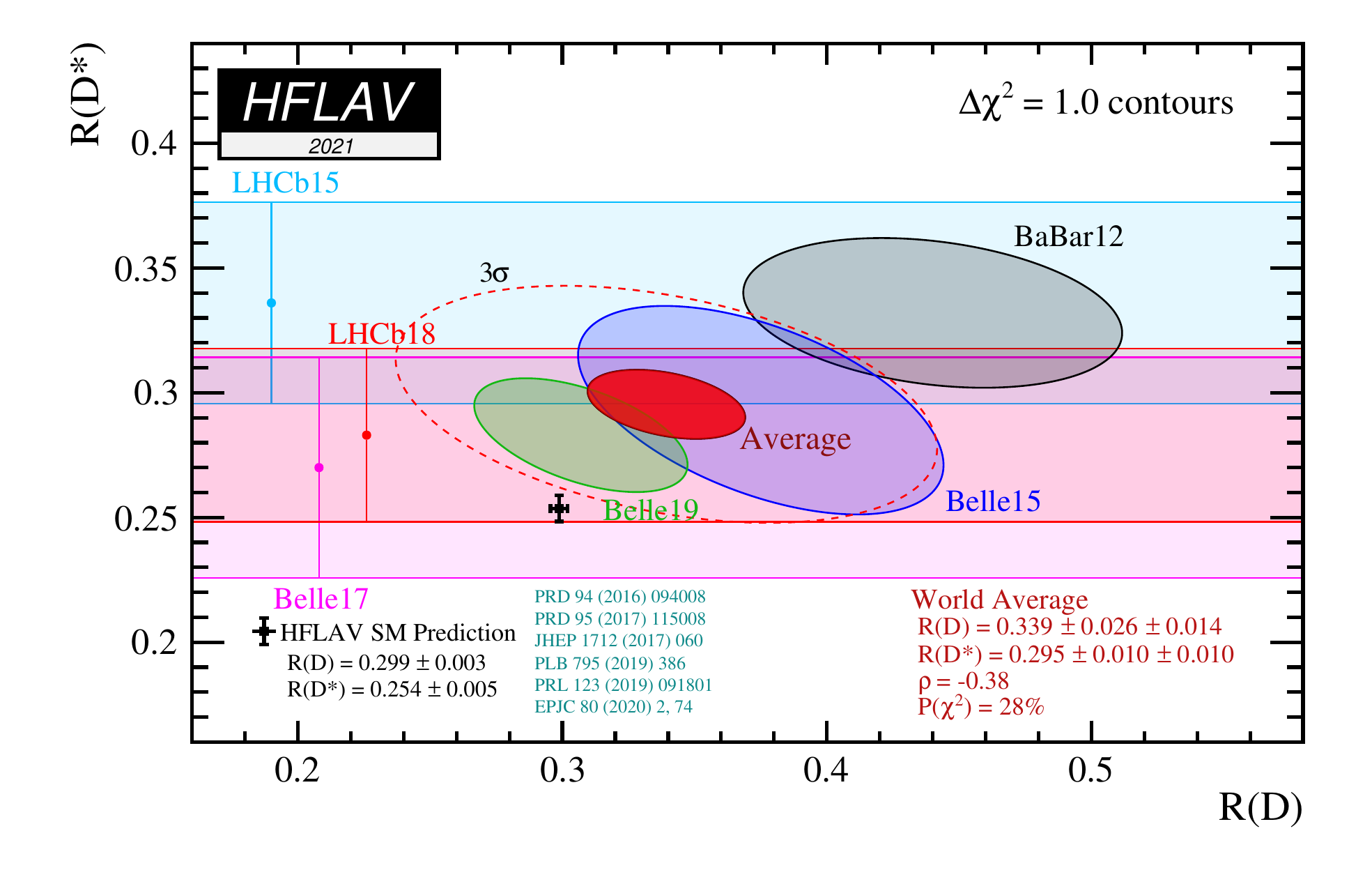}
\caption{The ratios of branching fractions $R(D)=\mathcal{B}(B\to D\tau^+\nu)/\mathcal{B}(B\to D\ell^+\nu)$ and $R(D^*)=\mathcal{B}(B\to D^*\tau^+\nu)/\mathcal{B}(B\to D^*\ell^-\nu)$, where $\ell$ denotes muons or electrons, are predicted precisely in the SM to be $R(D)=0.299\pm0.003$, $R(D^*)=0.254\pm0.005$ (the black point in this figure). The averages of experimental measurements of these ratios correspond to the red ellipse, which exceed the SM predictions with a combined significance of about $3.3\sigma$~\cite{HFLAV}.}\label{fig:RDRDst}
\end{figure}

The tree-level $b\to c\tau^-\nub$ decays are not rare but are nevertheless expected to be quite sensitive to physics beyond the SM as a result of the large $\tau$ lepton mass (for example, a charged Higgs boson would couple much more strongly to the $\tau$ than to the other leptons) \cite{LFUV-Overview-Whitepaper}. Experimental results are available for ratios of $b\to c\tau^-\nub$ to $b\to c\ell^-\nub$ branching fractions, typically denoted as $R(X_c)$ where $X_c$ is the charmed hadron in the final state. The world averages of experimental results for $R(D)$ and $R(D^*)$ exceed the SM predictions with a combined significance of $\sim 3\sigma$, again pointing to violation of lepton-flavor universality (Fig.~\ref{fig:RDRDst}).

The $b\to s\ell^+\ell^-$ and $b\to c\tau^-\nub$ anomalies have led to significant efforts by the HEP theory community to construct extensions of the SM that can explain them without introducing disagreements with other measurements. Several viable models have been identified as discussed further in Sec.~\ref{sec:BSM}. The models typically predict new particles with few-\tev masses, which could be discovered directly at future high-energy colliders (see, \eg, Refs.~\cite{Altmannshofer:2022xri,Azatov:2022itm,EF08-Report}). In addition, 
effective-field-theory arguments and specific models suggest possible correlations with NP signals in other low-energy observables, such as $K$ and $\tau$ decays \cite{Bordone:2017anc,Bordone:2017lsy,Fajfer:2018bfj,Borsato:2018tcz,Descotes-Genon:2020buf}.

\begin{figure}
\centering
\framebox{\includegraphics[width=0.95\linewidth]{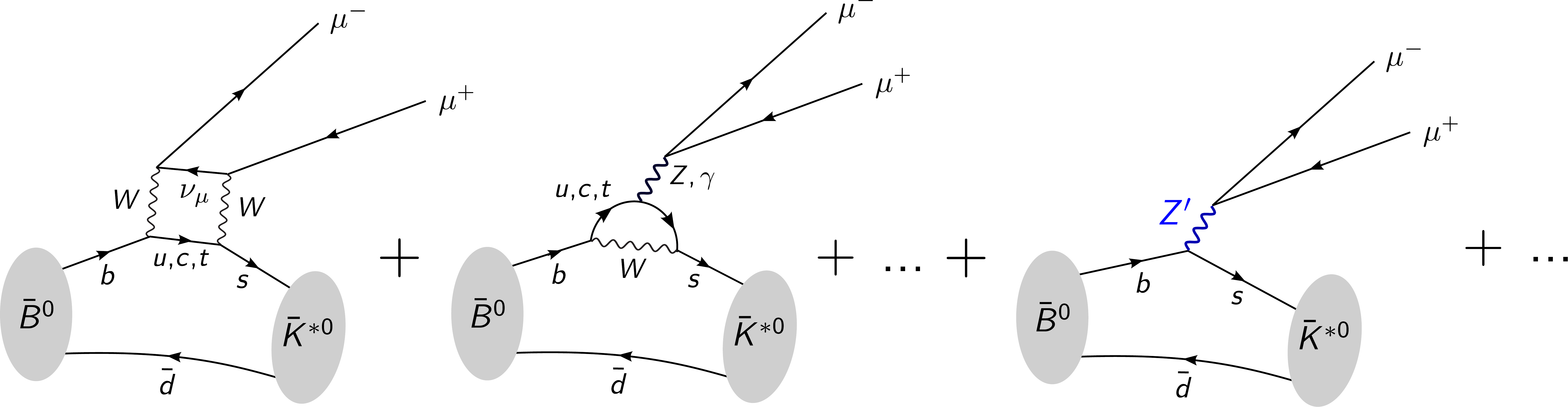}}\\[4ex]
\framebox{\includegraphics[width=0.95\linewidth]{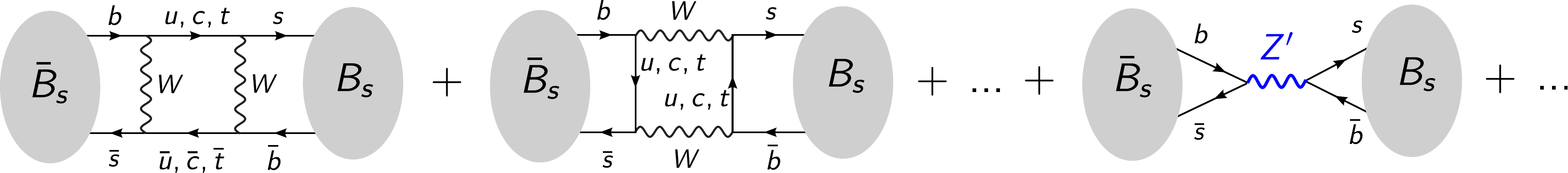}}\\[2ex]
\caption{Schematic representation of the (top) $\Bzb\to\Kbar{}^{*0}\mu^+\mu^-$ decay and (bottom) \Bs-\Bsb mixing amplitudes as sums over all possible Feynman diagrams. The diagrams on the left are examples of SM contributions, while the diagram on the right is an example of an NP contribution in theories with a flavor-changing neutral gauge boson $Z^\prime$.}
\label{fig:BtoKstarWZprime}\label{fig:BSmixingWZprime}
\end{figure}

\begin{figure}[t]
\centering
\includegraphics[width=0.5\textwidth]{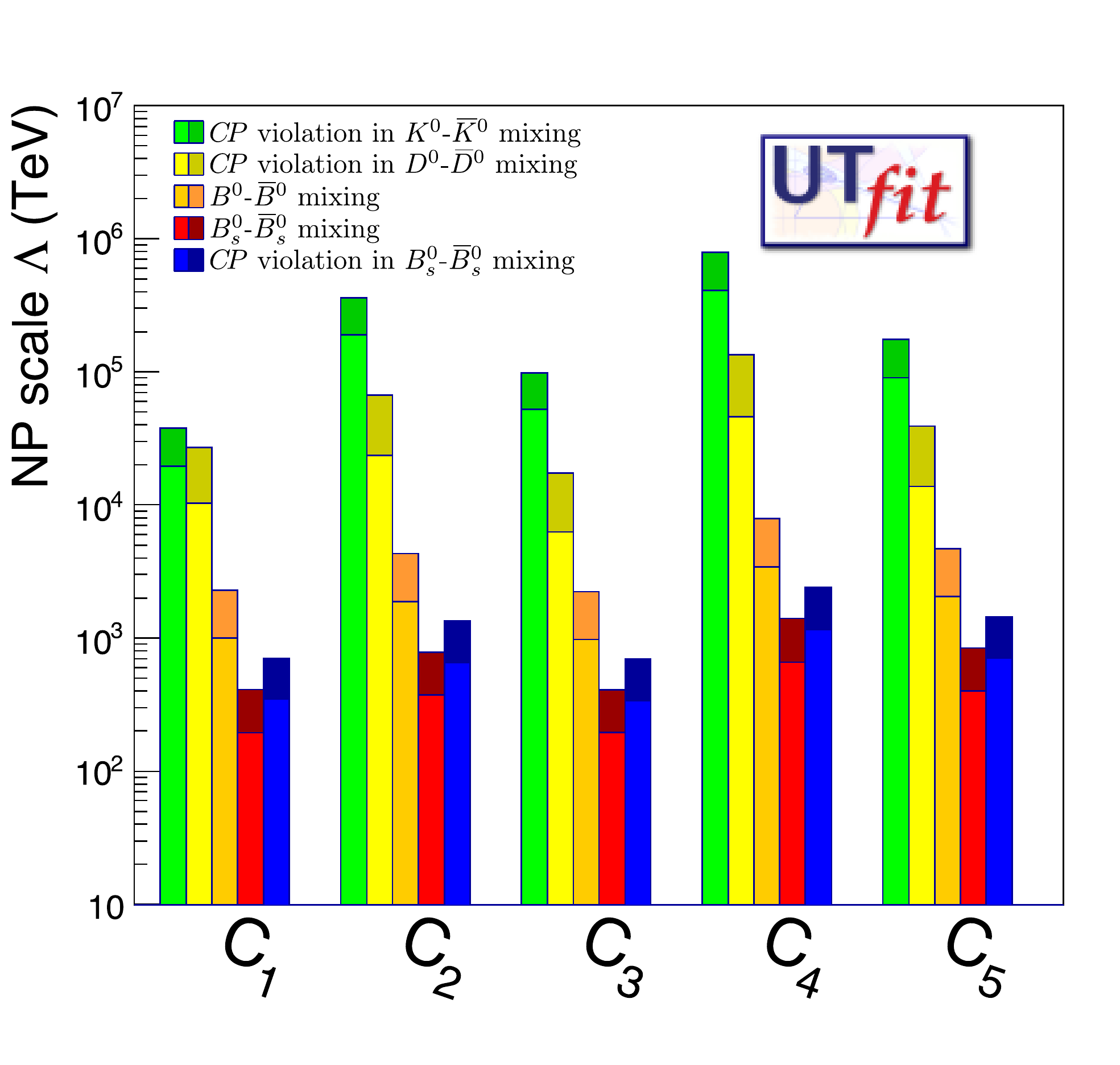}\hfil
\includegraphics[width=0.5\textwidth]{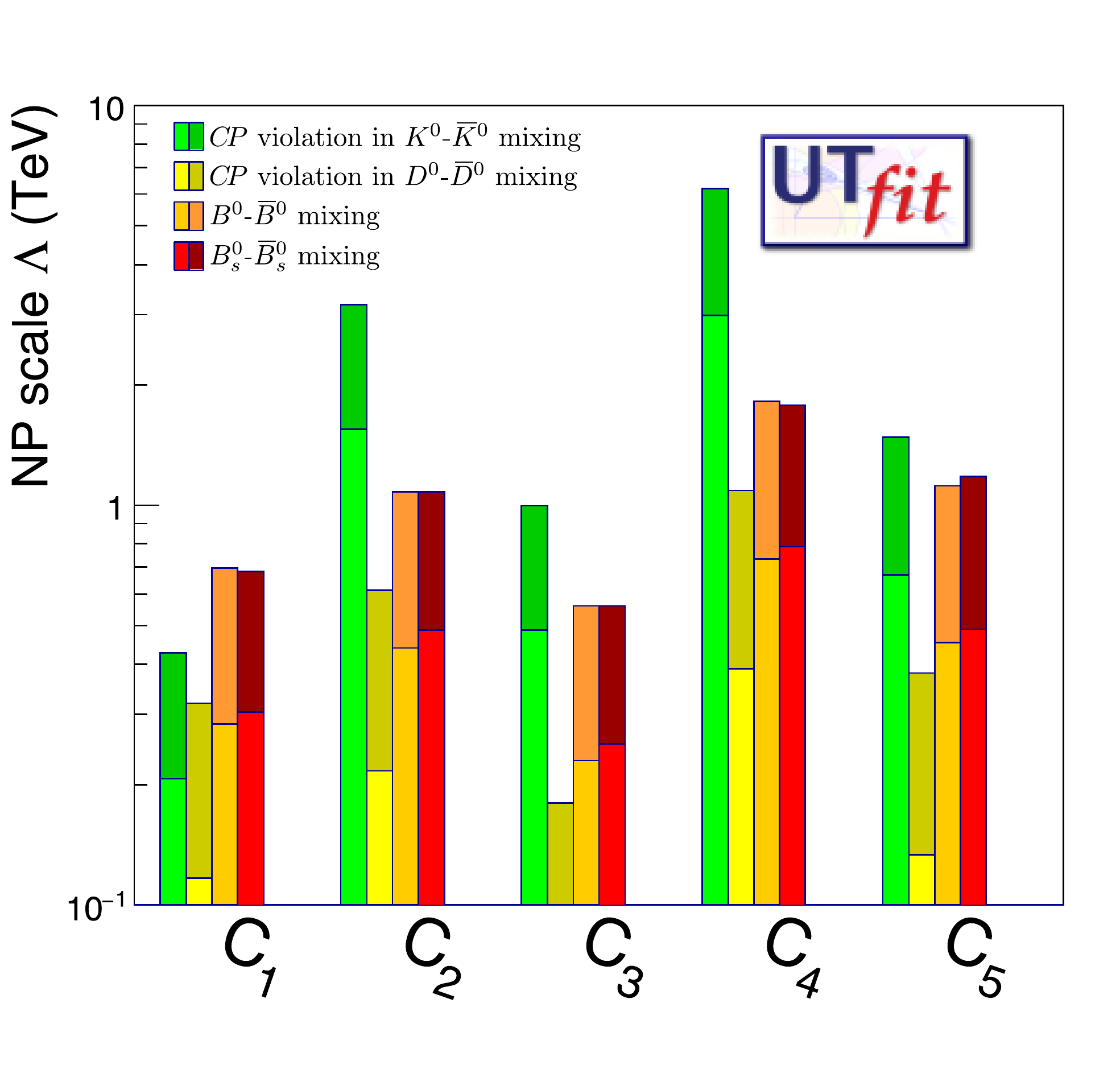}\\
\caption{Present (lighter) and future (darker) lower bounds at 95\% confidence level on the NP scale $\Lambda$ from $\Delta F=2$ transitions~\cite{UTfit:2007eik,Cerri:2018ypt}. The Wilson coefficients $C_i=F_iL_i/\Lambda^2$ ($i=1,...,5$) are the coupling of the NP dimension-six operators governing the $\Delta F=2$ transition: $Q_1^{q_iq_j}=(\bar{q}_{jL}^\alpha\gamma_\mu q_{iL}^\alpha)(\bar{q}_{jL}^\beta\gamma^\mu q_{iL}^\beta)$, $Q_2^{q_iq_j}=(\bar{q}_{jR}^\alpha q_{iL}^\alpha)(\bar{q}_{jR}^\beta q_{iL}^\beta)$, $Q_3^{q_iq_j}=(\bar{q}_{jR}^\alpha q_{iL}^\beta)(\bar{q}_{jR}^\beta q_{iL}^\alpha)$, $Q_4^{q_iq_j}=(\bar{q}_{jR}^\alpha q_{iL}^\alpha)(\bar{q}_{jL}^\beta q_{iR}^\beta)$, and $Q_5^{q_iq_j}=(\bar{q}_{jR}^\alpha q_{iL}^\beta)(\bar{q}_{jL}^\beta q_{iR}^\alpha)$ (with $\alpha$ and $\beta$ being color indices). On the left, NP is assumed to have arbitrary flavor structure ($F_i=1$) and to be strongly coupled with no loop suppression ($L_i=1$); on the right, NP is assumed to have minimal-flavor-violation couplings ($F_i=V_{\rm CKM}$) and to enter at one loop with weak coupling ($L_i=\alpha_2^2$, with $\alpha_2^2$ being the weak structure constant). The future bounds are based on expected sensitivities at Belle II (50\invab), BESIII, LHCb Upgrade II (300\invfb), and ATLAS/CMS (3\invab), and on improved theory inputs.\label{fig:NPscales}}
\end{figure}

The above anomalies are just an example -- though currently very intriguing -- of how dynamics beyond the SM could be discovered through measurements of heavy-quark transitions. However, to maximize the discovery potential, and to discern the type of NP, it is essential to investigate many different observables. Typically, any given NP model predicts correlated effects in several observables. Examples are illustrated in Fig.~\ref{fig:BtoKstarWZprime} for the cases of the rare decay $\Bz\to K^{*0}\mu^+\mu^-$ and of the \Bs-\Bsb mixing amplitude in theories with a flavor-changing neutral gauge boson $Z^\prime$. Such theories have been proposed as explanations of the deviations observed in $b\to s\mu^+\mu^-$ observables, and can be constrained further by comparing theory and experiment for the \Bs-\Bsb oscillation frequency~\cite{DiLuzio:2018wch,Charles:2020dfl}. In fact, meson-antimeson mixing provides to date the most stringent constraints on baryon- and lepton-number conserving NP, reaching energy scales of $O(10^5)$\tev for strongly-coupled NP with arbitrary flavor structure, as shown in the left panel of Fig.~\ref{fig:NPscales}. This extraordinary reach is due both to the Glashow-Iliopoulos-Maiani (GIM) mechanism and to the hierarchical structures of the quark masses and of the Cabibbo-Kobayashi-Maskawa (CKM) matrix. These bounds are clearly beyond the reach of any direct-detection experiment and strongly suggest us that any NP close to the electroweak scale must have a hierarchical flavor structure analogous to that of the SM. Indeed, the bound on the NP scale can be lowered to a few \tev (right panel of Fig.~\ref{fig:NPscales}) by requiring NP to have minimal flavor violation, \ie, the absence of new sources of flavor violation beyond the SM Yukawa couplings (\eg, as in composite Higgs models). In the minimal-flavor-violation case, the sensitivity becomes comparable to, and complements, other indirect probes of NP such as electroweak precision observables or Higgs couplings (see, \eg, Refs.~\cite{deBlas:2016ojx,Ellis:2018gqa,EF-Report}).

The coming two decades will not bring a substantial increase in the energy scales directly probed at colliders~\cite{EF-Report}. However, as discussed in the following, a remarkable increase in precision is expected for many heavy-quark observables. Weak decays of $b$ and $c$ quarks then offer a unique opportunity to reveal new phenomena, and/or strongly shape our expectations for beyond-SM dynamics, before the next energy-frontier machine will become available (see, \eg, the expected impact on the bounds on the NP scale from $\Delta F=2$ transitions in Fig.~\ref{fig:NPscales}).

\section{Experimental efforts in the next two decades\label{sec:experiments}}
\begin{figure}[t]
\centering
\includegraphics[width=\textwidth,trim={340 210 260 330},clip=true]{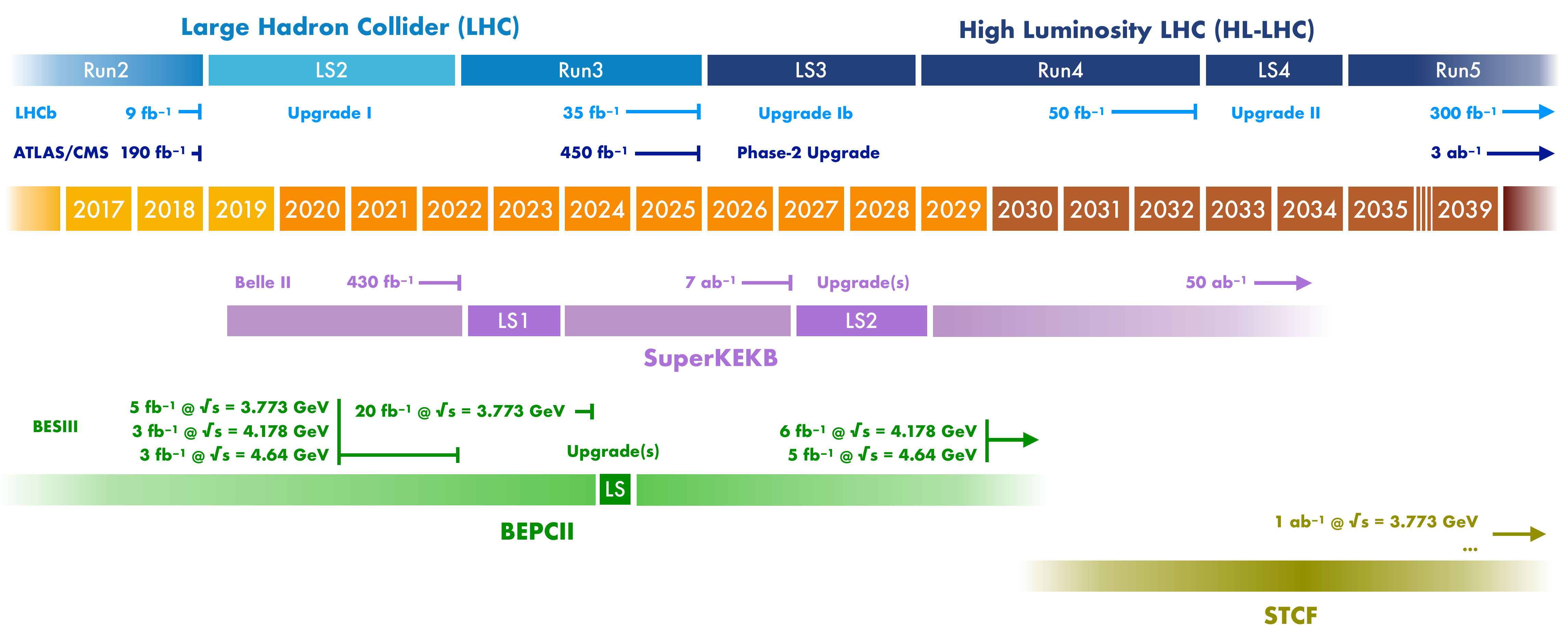}
\caption{Timeline of planned/proposed experiments in the next two decades~\cite{LHCSchedule,SuperKEKBSchedule,BESIII:2020nme,Lyu:2021tlb}.\label{fig:timeline}}
\end{figure}

Several experiments with beauty- and charm-quark physics in their programs are in operation or planned for the 2020s (Fig.~\ref{fig:timeline}). The Belle II experiment at the SuperKEKB asymmetric $e^+e^-$ collider is a major improvement over its predecessors Belle and BaBar. The experiment has been running since 2019 and, until SuperKEKB Long Shutdown 1 in Summer 2022, has collected $\sim430\invfb$ of integrated luminosity -- corresponding to roughly the sample size collected by BaBar. During the same period SuperKEKB has achieved a record peak luminosity of $4.7\times10^{34}\cm^{-2}\sec^{-1}$. The experiment will produce a variety of world-leading results continuously as it proceeds towards the goal of collecting an integrated luminosity of 50\invab by the mid-2030s~\cite{BelleII-Whitepaper,SuperKEKBSchedule}. To achieve this goal, SuperKEKB needs to reach a peak luminosity of $6.5\times10^{35}\cm^{-2}\sec^{-1}$ and be upgraded during Long Shutdown 2, which is currently scheduled for 2027-2028~\cite{Forti:2022mti,Endo:2022imj,SuperKEKBSchedule}. An international task force has been formed to provide advice to SuperKEKB on the possible upgrade options, which include a redesign of the interaction region and of the final focusing system. Long Shutdown 2 provides the possibility to upgrade parts of the Belle II detector as well. A new vertex detector might be required to accommodate the new interaction-region design, and other sub-detectors might require improved robustness against increasing machine background~\cite{Forti:2022mti,Natochii:2022vcs}.

The LHCb experiment, after having collected about 9\invfb of $pp$ collisions during Run 1 and 2 of the LHC, has just started the operation of its first upgrade during Run 3. By 2032, LHCb Upgrade I expects to collect a sample corresponding to an integrated luminosity of 50\invfb~\cite{LHCb-Whitepaper}. Considering the increased collision energy and the enhanced online-selection efficiency, the yield of beauty and charm hadrons available for analyses will increase by factors of 5 to 10, depending on the final states, compared to the currently available data from Run 1 and Run 2. On the same timescale, ATLAS and CMS will continue to contribute significantly in some selected areas, such as in decays with dimuons in the final state~\cite{AtlasCMS-Whitepaper}. For the HL-LHC both experiments are planning significant modification of the detectors (Phase-2 upgrades scheduled during Long Shutdown 3 in 2026-2028) to maintain effective data taking and event reconstruction at increased luminosity and pileup. Particularly relevant for heavy-flavor physics are upgrades to the tracking systems, which would result in improved mass and decay-time resolutions, and to the trigger systems, to maintain the online selection efficient at the relatively low transverse momenta typical of the final state muons from beauty decays.

Finally, the BESIII experiment at BEPCII uses $e^+e^-$ collisions with center-of-mass energies ranging from 2 to 5\gev to study the broad spectrum of physics accessible in the $\tau$-charm energy region~\cite{BESIII-Whitepaper}. Since the start of operations in 2009, BESIII has collected more than 35\invfb of data, comprising several data samples that are particularly useful for studying weak decays of charm hadrons, such as 5\invfb of $\psi(3770)\to\Dz\Dzb$ data, 3\invfb at $\sqrt{s} = 4.178$ (near the \mbox{$\Dsp\!\Dssm$} threshold), and more than 3\invfb at $\sqrt{s}=4.64\gev$ (above the \Lc\Lcb threshold). The experiment will run at least for the next 5-10 years, during which new upgrades for both the detector and accelerator are being considered. In particular, BEPCII upgrades aim to first increase the maximum collision energy to 5.6\gev and then to increase the peak luminosity by a factor of three (for collision energies above 4\gev). The goal is for BESIII to integrate 20\invfb at $\sqrt{s}=3.773\gev$ (before the scheduled BEPCII upgrade in June 2024), $6\invfb$ at $\sqrt{s} = 4.178$, and 5\invfb at $\sqrt{s}=4.64\gev$~\cite{BESIII:2020nme}.

These experimental efforts will complement one another, making possible a wide range of precision measurements that would be unfeasible at a single facility. Since most of the current heavy-flavor results are severely limited by their statistical precision, the ability to access much larger samples of $b$- and $c$-hadron decays is crucial. In this respect, the production rate of heavy flavored hadrons in $pp$ collisions gives LHC experiments a clear advantage with respect to the beauty and charm ``factories'' operating at $e^+e^-$ colliders. However, such advantage is mostly exploited in final states made of only charged particles thanks to the excellent tracking and vertexing detectors that, by precisely measuring their properties, allow to discriminate signal particles from backgrounds. In addition, all species of $b$ hadrons are produced at the LHC, including bottom-strange mesons and bottom baryons, which are kinematically forbidden at Belle II. Belle II and BESIII have unique capabilities that give them advantages over hadron-collider experiments despite the lower production rates. Reconstruction of neutral particles (photons and neutral pions) is nearly as efficient and precise as that of charged particles. Because the initial state is known and the detectors are nearly hermetic, fully-inclusive final states and reconstruction of particles with no direct signature in the detector (such as neutrinos and \KL mesons) becomes accessible. Reconstruction efficiencies at beauty and charm factories are largely uniform as a function of the decay kinematics, which offers an advantage in measurements that involve multibody decays. Coherent \mbox{$\Bz\!\Bzb$} pair production at Belle II makes possible efficient determination of the flavor of the neutral $B$ meson at production (flavor tagging), which is key for several time-dependent \CP-violation measurements. Coherent \mbox{$\Dz\!\Dzb$} pair production through the $\psi(3770)\to\Dz\Dzb$ process at BESIII enables the measurement of quantum-correlated observables that cannot be accessed elsewhere. Examples are strong-phase differences between the \Dz and \Dzb amplitudes, which are important inputs for the precise determination of the CKM angle $\gamma$ and of the charm-mixing parameters in a model-independent fashion at Belle II and LHCb~\cite{Giri:2003ty,Bondar:2005ki,Bondar:2008hh,Bondar:2010qs,DiCanto:2018tsd,Poluektov:2017zxp}.

The HL-LHC will extend the LHC program through the 2030s and the first half of the 2040s. While no major upgrades are yet planned for ATLAS and CMS during this period, LHCb proposes to upgrade the entire detector to be able to run at an instantaneous luminosity of around $1.5\times10^{34}\cm^{-2}\sec^{-1}$, and collect a total of 300\invfb by the end of Run 6 (in the early 2040s)~\cite{LHCb-Whitepaper,LHCb-TDR-023}. The proposal is to install the new detector (Upgrade II) during LHC Long Shutdown 4 (2033-2034), with some preparatory work (Upgrade Ib) to be performed already during Long Shutdown 3 (2026-2028). The Upgrade Ib will also have benefits for the physics performance during Run 4, beyond what has been projected for LHCb Upgrade I. The challenge for the LHCb Upgrade II resides mostly in maintaining and extending the strengths of the LHCb Upgrade I detector, including its flexible software trigger, in the much harsher environment resulting from $\sim40$ interactions per bunch crossings. The detector must sustain radiation doses of up to 400\,MRad per year, be highly segmented to cope with large occupancy, and integrate timing information (with tens of \ps resolution) in the readout to be able to associate hits with the right primary interaction. These challenges require the development of novel technologies, some of which will likely be deployed in future HEP experiments~\cite{IF-Report}.

At Belle II, studies have started to explore upgrades beyond the currently planned program, such as beam polarization and ultra-high luminosity. The beam-polarization upgrade offers
unique and powerful sensitivities to NP via precision measurements of neutral-current couplings at 10\gev, and via studies of $\tau$-lepton properties and decays~\cite{SuperKEKB-Polarization}. Accelerator upgrades to reach a peak luminosity in excess of $1\times10^{36}\cm^{-2}\sec^{-1}$ and collect $\sim250\invab$ of integrated luminosity have recently been discussed~\cite{Forti:2022mti}. If timely, such an upgrade may effectively complement the heavy-flavor program of the HL-LHC experiments. However, the feasibility from the accelerator perspective is still unclear, and so is the upgrade timeline.

With BESIII expected to end by around 2030, a Super $\tau$-Charm factory (STCF)~\cite{STCF-Whitepaper,Lyu:2021tlb} has been proposed in China to continue and extend the physics program with $e^+e^-$ collisions at energies between 2 and 7\gev and with peak luminosity of at least $5\times10^{34}\cm^{-2}\sec^{-1}$. The current schedule foresees the construction to occur between 2024 and 2030, and at least 10 years of operations. Upgrades to further increase the luminosity and for the implementation of a polarized $e^-$ beam are also proposed for 2041-2042, followed by an additional 5 years of data taking.

\subsection{Expected experimental progress on key observables\label{sec:exp-progress}}
In a short summary such as this report, it is impractical to discuss all interesting observables in $b$ and $c$ physics. Thus we focus on a small subset of observables that are currently of high interest, either because their measurements show possible deviations from SM predictions, or because they are limited by experimental uncertainties and therefore offer an opportunity for significant improvement in precision over the next 10-20 years. A summary of the experimental prospects for many of these key measurements is given in Tab.~\ref{tab:sensitivity}.

\begin{sidewaystable}[p]
\centering
\resizebox{\textwidth}{!}{
\begin{tabular}{lccccccccc}
\toprule
Observable & Current & \multicolumn{2}{c}{Belle II} & \multicolumn{2}{c}{LHCb} & ATLAS & CMS & BESIII & STCF\\
 & best & 50\invab & 250\invab & 50\invfb & 300\invfb & 3\invab & 3\invab & 20\invfb $(*)$ &  1\invab $(*)$\\
\midrule
\textbf{Lepton-flavor-universality tests} \\
$R_K(1<q^2<6\gevgevcccc)$ & 0.044~\cite{LHCb:2021trn} & 0.036 & 0.016 & 0.017 & 0.007 \\
$R_{K^*}(1<q^2<6\gevgevcccc)$ & 0.12~\cite{LHCb:2017avl} & 0.032 & 0.014 & 0.022 & 0.009 \\
$R(D)$ & 0.037~\cite{Belle:2019rba} & 0.008 & $<0.003$ & na & na \\
$R(D^*)$ & 0.018~\cite{Belle:2019rba} & 0.0045 & $<0.003$ & 0.005 & 0.002 \\
\midrule
\textbf{Rare decays} \\
$\BF(\Bs\to\mu^+\mu^-)$ [$10^{-9}$] & 0.46~\cite{LHCb:2021vsc,LHCb:2021awg} & & & na & 0.16 & 0.46--0.55 & 0.39\\
$\BF(\Bz\to\mu^+\mu^-)/\BF(\Bs\to\mu^+\mu^-)$ & 0.69~\cite{LHCb:2021vsc,LHCb:2021awg} & & & 0.27 & 0.11 & na & 0.21 \\
$\BF(\Bz \to K^{*0} \tau^+\tau^-)$ UL [$10^{-3}$] & 2.0~\cite{BaBar:2016wgb,Belle:2021ndr} & 0.5 & na \\
$\BF/\BF_{\rm SM}(B^+ \to K^+ \nu\nub)$ & 1.4~\cite{BaBar:2013npw,Belle:2017oht} & 0.08--0.11 & na \\
$\BF(B\to X_s\gamma)$ & 10\%~\cite{BaBar:2012fqh,Belle:2014nmp} & 2--4\% & na \\
\midrule
\textbf{CKM tests and \CP violation} \\
$\alpha$ & 5\degrees~\cite{BaBar:2014omp} & 0.6\degrees & 0.3\degrees \\
$\sin2\beta(\Bz\to\jpsi\KS)$ & 0.029~\cite{Belle:2012paq} & 0.005 & 0.002 & 0.006 & 0.003 \\
$\gamma$ & 4\degrees~\cite{LHCb:2021dcr} & 1.5\degrees & 0.8\degrees & 1\degrees & 0.35\degrees & & & $0.4\degrees\,(\dagger)$ & $<0.1\degrees\,(\dagger)$\\
$\phi_s(\Bs\to\jpsi\phi)$ & 32\mrad~\cite{LHCb:2019nin} & & & 10\mrad & 4\mrad & 4--9\mrad & 5--6\mrad \\
$|V_{ub}|(\Bz\to\pi^-\ell^+\nu)$ & 5\%~\cite{BaBar:2010efp,Belle:2010hep} & 2\% & $<1\%$ & na & na \\
$|V_{ub}|/|V_{cb}|(\Lb\to p\mu^-\nub)$ & 6\%~\cite{LHCb:2015eia} & & & 2\% & 1\% \\
$f_{D^+}|V_{cd}|(D^+\to\mu^+\nu)$ & 2.6\%~\cite{BESIII:2013iro} & 1.4\% & na & & & & & 1.0\% & 0.15\%\\
$\SCP(\Bz\to\eta^\prime\KS)$ & 0.08~\cite{BaBar:2008ucf,Belle:2014atq} & 0.015 & 0.007 & na & na \\
$\ACP(\Bz\to\KS\pi^0)$ & 0.15~\cite{BaBar:2008ucf,Belle:2008kbm} & 0.025 & 0.018 & na & na \\
$\ACP(D^+\to\pi^+\pi^0)$ & $11\times10^{-3}$~\cite{LHCb:2021rou} & $1.7\times10^{-3}$ & na & na & na & & & na & na\\
$\Delta x(\Dz\to\KS\pi^+\pi^-)$ & $18\times10^{-5}$~\cite{LHCb:2021ykz} & na & na & $4.1\times10^{-5}$ & $1.6\times10^{-5}$ & & & & \\
$A_\Gamma(\Dz\to K^+K^-,\pi^+\pi^-)$ & $11\times10^{-5}$~\cite{LHCb:2021vmn} & na & na & $3.2\times10^{-5}$ & $1.2\times10^{-5}$ & & & & \\
\bottomrule
\end{tabular}}
\caption{Projected uncertainties (or 90\% CL upper limits) in several key heavy-flavor observables over the next two decades. A missing entry means that the observable cannot be measured, the abbreviation {\rm na} means that, although the observable can be measured, the projected uncertainty is not available. Projections are taken from Refs.~\cite{BelleII-Whitepaper,Forti:2022mti,Belle-II:2018jsg} (Belle~II), Refs.~\cite{LHCb:2018roe,LHCb-TDR-023} (LHCb), Ref.~\cite{AtlasCMS-Whitepaper} (ATLAS and CMS), Refs.~\cite{BESIII:2020nme,STCF-Whitepaper} (BESIII and STCF). $(*)$ Integrated luminosity at $\sqrt{s}=3.773$. $(\dagger)$ Projected uncertainties on $\gamma$ resulting from BESIII/STCF measurements of the $D$ strong-phase differences, which will contribute as external inputs to the Belle~II and LHCb measurements.\label{tab:sensitivity}}
\end{sidewaystable}

\subsubsection{Lepton-flavor-universality tests}
\begin{figure}[t]
\centering
\includegraphics[height=0.28\textheight]{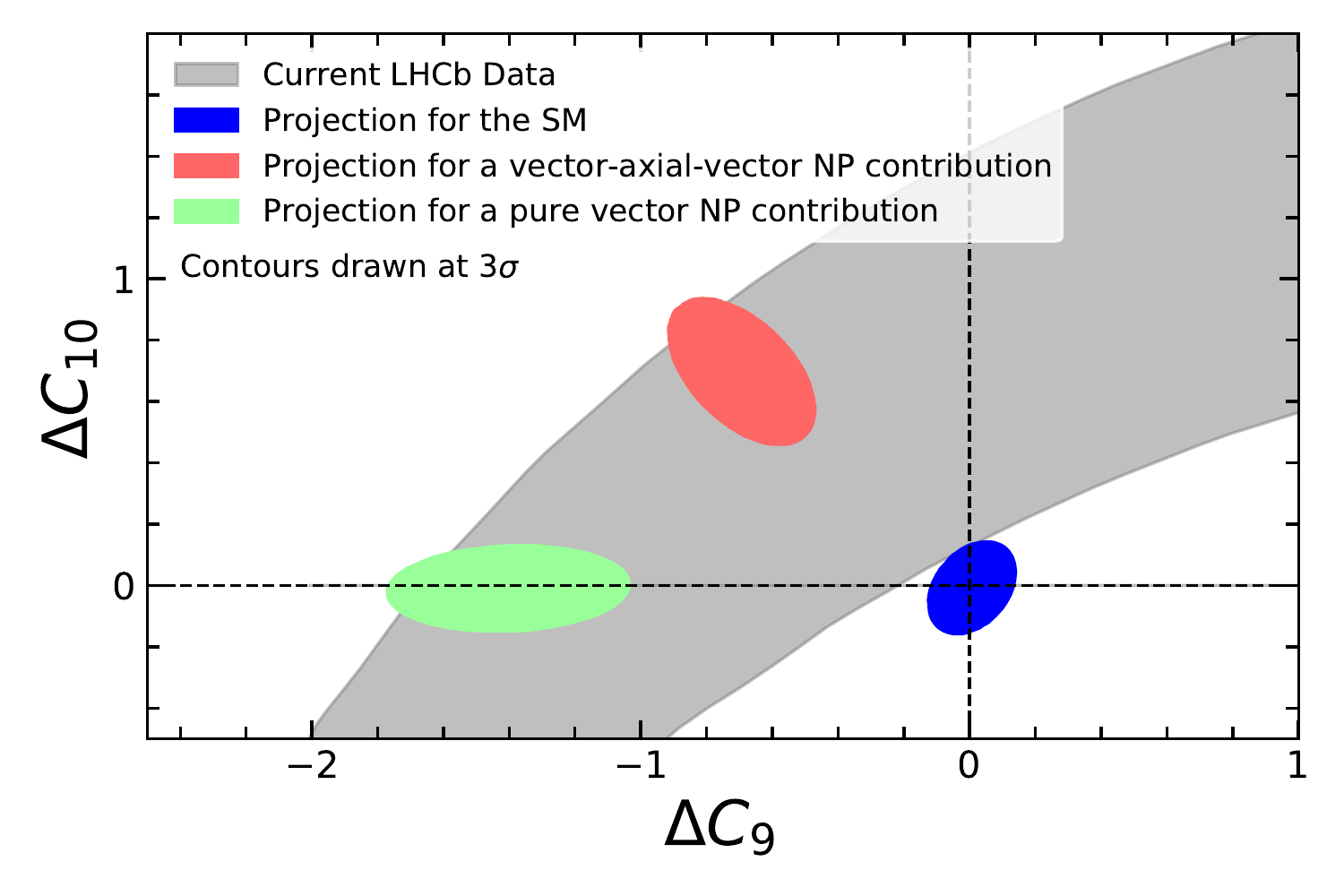}\hfil
\includegraphics[height=0.28\textheight]{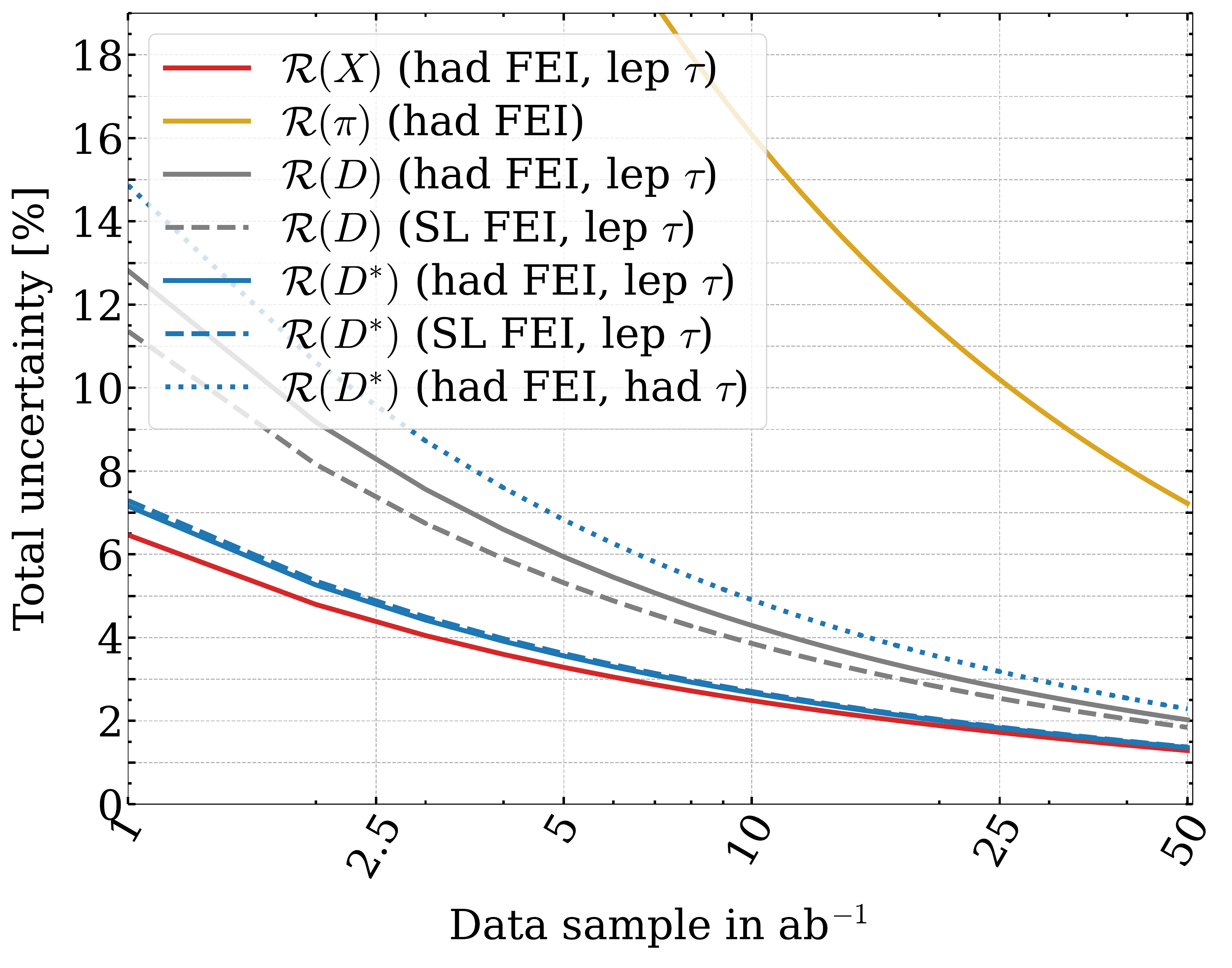}\\
\caption{(Left) Projected sensitivity at LHCb to the difference between muon- and electron-mode contributions to the $C_9$ and $C_{10}$ couplings in different scenarios~\cite{LHCb-TDR-023}. Blue, red and green filled regions show the $3\sigma$ uncertainty contours under each scenario with the LHCb Upgrade II data set. The grey region shows the current $3\sigma$ uncertainty. (Right) Expected Belle~II sensitivity for various tests of lepton-flavor universality in semitauonic $B$ decays as a function of integrated luminosity~\cite{BelleII-Whitepaper}. The FEI acronym refers to the algorithm for reconstruction of the partner $B$-meson~\cite{Keck:2018lcd}.\label{fig:Belle2-RDx}\label{fig:LHCb-c9c10}}
\end{figure}

The next decade should clarify the hints of lepton-flavor-universality violation observed in recent years thanks to the large data samples expected at LHCb Upgrade I and Belle II. Measurements of LFU observables in $b\to s\ell^+\ell^-$ decays will reach 1\%-level uncertainties, a precision sufficient to establish or reject the level of LFU violation seen in the current measurements. The LHCb Upgrade II data set will then open new avenues with sensitivity to even cleaner observables, such as the difference between the values of $C_9$ and $C_{10}$ for $b\to s e^+e^-$ and $b\to s\mu^+\mu^-$ transitions, through the measurements of angular distributions~\cite{Mauri:2018vbg,Sibidanov:2022gvb}. The achievable precision will be crucial to distinguish between different NP models (Fig.~\ref{fig:LHCb-c9c10}). Additionally, the LHCb Upgrade II sample will allow lepton-flavor-universality tests in the related, and further suppressed, $b\to d\ell^+\ell^-$ transitions, which would further constrain the dynamics. For example, the statistical precision on the ratio $\BF(B^+\to\pi^+\mu^+\mu^-)/\BF(B^+\to\pi^+ e^+e^-)$ is expected to reach a few percent.

Tests of lepton-flavor universality in $\Bz\to\DorDsm\tau^+\nu$ decays are expected to be dominated by Belle~II, thanks to the ability to constrain the kinematics of the undetected neutrinos by utilizing the precise knowledge of the \mbox{$\Bz\!\Bzb$}-pair production mechanism. LHCb will also contribute, in particular, by performing measurements of semitauonic rates of other $b$ hadrons not accessible at Belle II, such as $\Bs\to\DsorDssm\tau^+\nu$, $B_c^+\to\jpsi\tau^+\nu$ and $\Lb\to\Lc\tau^-\nub$. Measurements of observables related to angular distributions, such as the $\tau^+$ and $D^{*-}$ polarization fractions, will provide supplementary sensitivity to non-SM physics and key information to decipher the dynamics (see, \eg, Ref.~\cite{Duraisamy:2013pia,Ligeti:2016npd,Hill:2019zja,Bhattacharya:2020lfm,Bernlochner:2020tfi}). Furthermore, Belle~II has the unique ability to measure the inclusive ratio $R(X) = \BF(B\to X\tau^+\nu)/\BF(B\to X\ell^+\nu)$ -- where $X$ is any system made of one or more hadrons -- whose phenomenological interpretation is based on different theory inputs compared to the exclusive observables, and will perform, for the first time, measurements of $b\to u\tau^-\nub$ decays. As shown in Fig.~\ref{fig:Belle2-RDx}, Belle II will achieve $\mathcal{O}(1\%)$ sensitivities on most quantities using 50\invab of integrated luminosity.

Other possibilities to test electron \vs muon universality in semileptonic charm and beauty decays have also been recently proposed, with many having good prospects at BESIII, STCF and Belle II~\cite{BESIII-Whitepaper,STCF-Whitepaper,Bobeth:2021lya,Bhattacharya:2022cna,Bhattacharya:2022bdk}. One example is the difference between the lepton forward-backward asymmetries for $B^0\to D^{*-}\mu^+\nu$ and $B^0\to D^{*-}e^+\nu$ decays~\cite{Bobeth:2021lya,Bhattacharya:2022cna,Bhattacharya:2022bdk}. This difference can be precisely measured at Belle II.

\subsubsection{Rare decays}
The purely leptonic rare decay $\Bs\to\mu^+\mu^-$ has very small branching fraction in the SM and, as such, it has been historically considered one of the ``golden'' channels for flavor-changing neutral-current $b$-hadron decays. The average of the experimental measurements of $\BF(\Bs\to\mu^+\mu^-)$, $(3.01\pm0.35)\times10^{-9}$, is dominated by statistical uncertainties that are larger than the theory uncertainties on the predicted SM value, $(3.66\pm0.14)\times10^{-9}$~\cite{Beneke:2019slt}. The experimental precision is expected to closely approach the SM uncertainty with the LHCb Upgrade II data set and to match it when combining with ATLAS and CMS 3\invab results (Tab.~\ref{tab:sensitivity}). A related probe for beyond-SM dynamics, which is particularly powerful in constraining supersymmetry or models with minimal flavor violation, is the ratio between $\Bz\to\mu^+\mu^-$ and $\Bs\to\mu^+\mu^-$ branching fractions. This will be measured with better than 10\% relative precision by the combination of the HL-LHC measurements. Thanks to its superior vertexing and flavor-tagging capabilities as compared to ATLAS and CMS, LHCb Upgrade II will furthermore have the unique ability to measure the \CP-violation parameters $A_{\Delta\Gamma}$ and $S_{\CP}$ with a time-dependent analysis of $\Bs\to\mu^+\mu^-$ decays~\cite{LHCb-TDR-023}. These are important observables that, if different from their SM expectations of unity and zero respectively, would provide unambiguous evidence for new dynamics~\cite{DeBruyn:2012wk}.

Besides from lepton-flavor-violation tests, differential decay rates of semileptonic $b\to s\ell^+\ell^-$ decays as a function of the squared dilepton mass, $q^2$, and of angular observables provide a wealth of information to constrain all Wilson coefficients in the effective Hamiltonian. Hints of NP in $C_9$ firstly arose from measurements of the $q^2$-dependent $\Bz\to K^{*0}\mu^+\mu^-$ and $\Bs\to\phi\mu^+\mu^-$ angular distributions and branching ratios at LHCb~\cite{LHCb:2013ghj,LHCb:2013zuf,LHCb:2013tgx,Descotes-Genon:2013wba,Horgan:2013pva,LHCb:2021zwz,LHCb:2020lmf}. Experimental progress in this area is expected to be dominated by LHCb, with contributions from Belle II and the Phase-2 upgrades of ATLAS and CMS~\cite{Rare-decay-Overview-Whitepaper}. The absence of charged leptons in the final state -- which removes theory uncertainties due to charm-loop effects~\cite{Khodjamirian:2010vf} -- makes $b\to s\nu\nub$ decays particularly interesting complementary probes of the non-SM physics scenarios proposed to explain the $b\to s\ell^+\ell^-$ anomalies~\cite{Browder:2021hbl}. Belle II will be the only experiment capable of exploring these key channels in the next decades. As an example, it has the potential to observe $B^+\to K^+\nu\nub$ decays at the SM rate with only 5\invab of integrated luminosity and severely constrain various non-SM extensions such as models with leptoquarks, axions, feebly interacting, or dark-matter particles~\cite{BelleII-Whitepaper}. Given that many NP models predict the largest effects for the 3rd generation, additional complementary information will arise from improved searches at LHCb and Belle II of $b\to s\tau^+\tau^-$ decays. For example, the limits on $\BF(\Bs\to\tau^+\tau^-)$ and $\BF(B^+\to K^+\tau^+\tau^-)$ are both expected to improve by an order of magnitude or more at the end of LHCb Upgrade II~\cite{Rare-decay-Overview-Whitepaper,Cornella:2020aoq}; and Belle II is expected to improve current limits on the $\Bz\to\tau^+\tau^-$ and $\Bz\to K^{*0}\tau^+\tau^-$ branching fractions by factors of 20 and 4, respectively, with 50\invab of integrated luminosity~\cite{BelleII-Whitepaper,Belle-II:2018jsg}.

LHCb and Belle II will access a wide range of $b\to s(d)\gamma$ decays. Belle II is expected to have the best sensitivity in the next decade to observables based on exclusive $B$ decays, reaching on many percent or sub-percent precision with 50\invab of integrated luminosity~\cite{BelleII-Whitepaper,Belle-II:2018jsg}. LHCb has unique access to time-dependent \CP-violation observables in $\Bs\to\phi\gamma$ decays, for which an improved calorimeter during Upgrade II would be critical to keep systematic uncertainties comparable or below the expected statistical uncertainties~\cite{LHCb-Whitepaper,LHCb-TDR-023}. Belle II will also study these transitions inclusively. The precise and reliable SM prediction of the inclusive $B\to X_s\gamma$ rate, where $X_s$ identifies a particle or system of particles with strangeness, makes it an extremely sensitive probe for beyond-SM dynamics~\cite{Misiak:2020vlo,Huber:2020vup}. In addition, the inclusive analysis enables the determination of observables like the $b$-quark mass and can provide input to determinations of $|V_{ub}|$~\cite{Belle-II:2018jsg}. Depending on the assumed detector performance in rejecting neutral hadrons faking photons, Belle II will reach a relative precision on $\BF(B\to X_s\gamma)$ between 2\% and 4\% with the 50\invab data set, which is comparable to the theory prediction. Moreover, it will have the ability to explore the photon-energy spectrum at much lower energies than before~\cite{BelleII-Whitepaper,Belle-II:2018jsg}.

Rare and forbidden decays of charm hadrons probe beyond-SM contributions in $c\to u$ transitions and are therefore complementary to searches done in the $b$ sector. Despite the SM rate being dominated by long-distance dynamics, the effective GIM cancellation and (approximate) symmetries of the charm system allow to define various null-test observables related to angular distributions and \CP violation, which have high discovery potential in the near future~\cite{Gisbert:2020vjx}. First measurements of such observables in $\Dz\to K^+K^-\mu^+\mu^-$ and $\Dz\to\pi^+\pi^-\mu^+\mu^-$ decays have been recently performed at LHCb~\cite{LHCb:2021yxk}. The experimental precision has already reached $0.1-0.01$ and, for the $\Dz\to K^+K^-\mu^+\mu^-$ results, the overall agreement with the SM is at the level of $2.7\sigma$. These measurements are expected to reach sub-percent precision during Upgrade II, and will be complemented by studies of other rare charm decays and by tests of lepton-flavor universality~\cite{LHCb:2018roe}. Lepton-flavor universality tests and unique studies of $c\to u\nu\nub$ decays are expected to be possible also at $e^+e^-$ colliders~\cite{Belle-II:2018jsg,BESIII-Whitepaper,STCF-Whitepaper}.

\subsubsection{Precise CKM-unitarity tests and new sources of \CP violation}\label{sec:CKMtests}
\begin{figure}[t]
\centering
\includegraphics[width=0.5\textwidth]{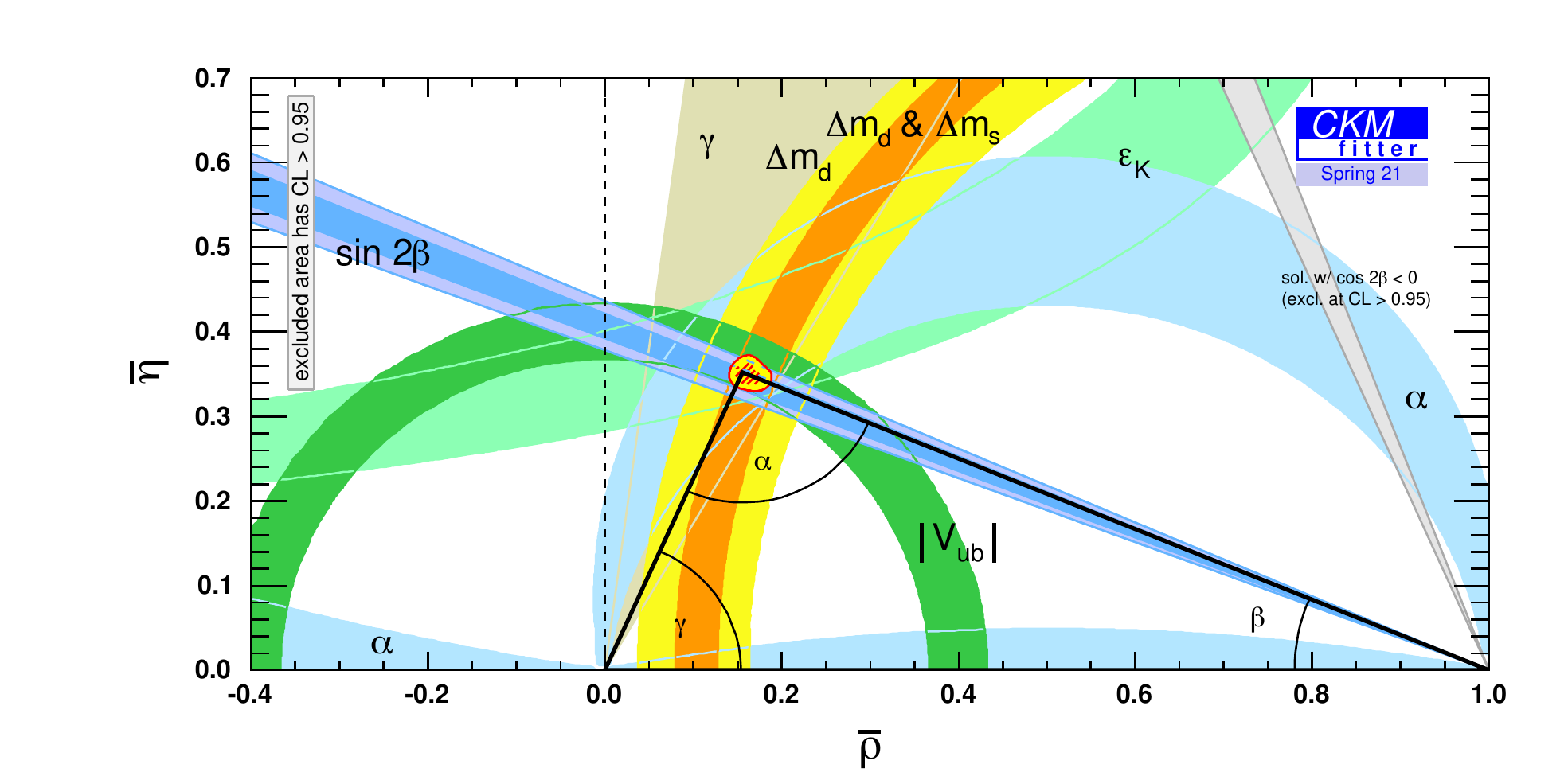}\hfil
\includegraphics[width=0.5\textwidth]{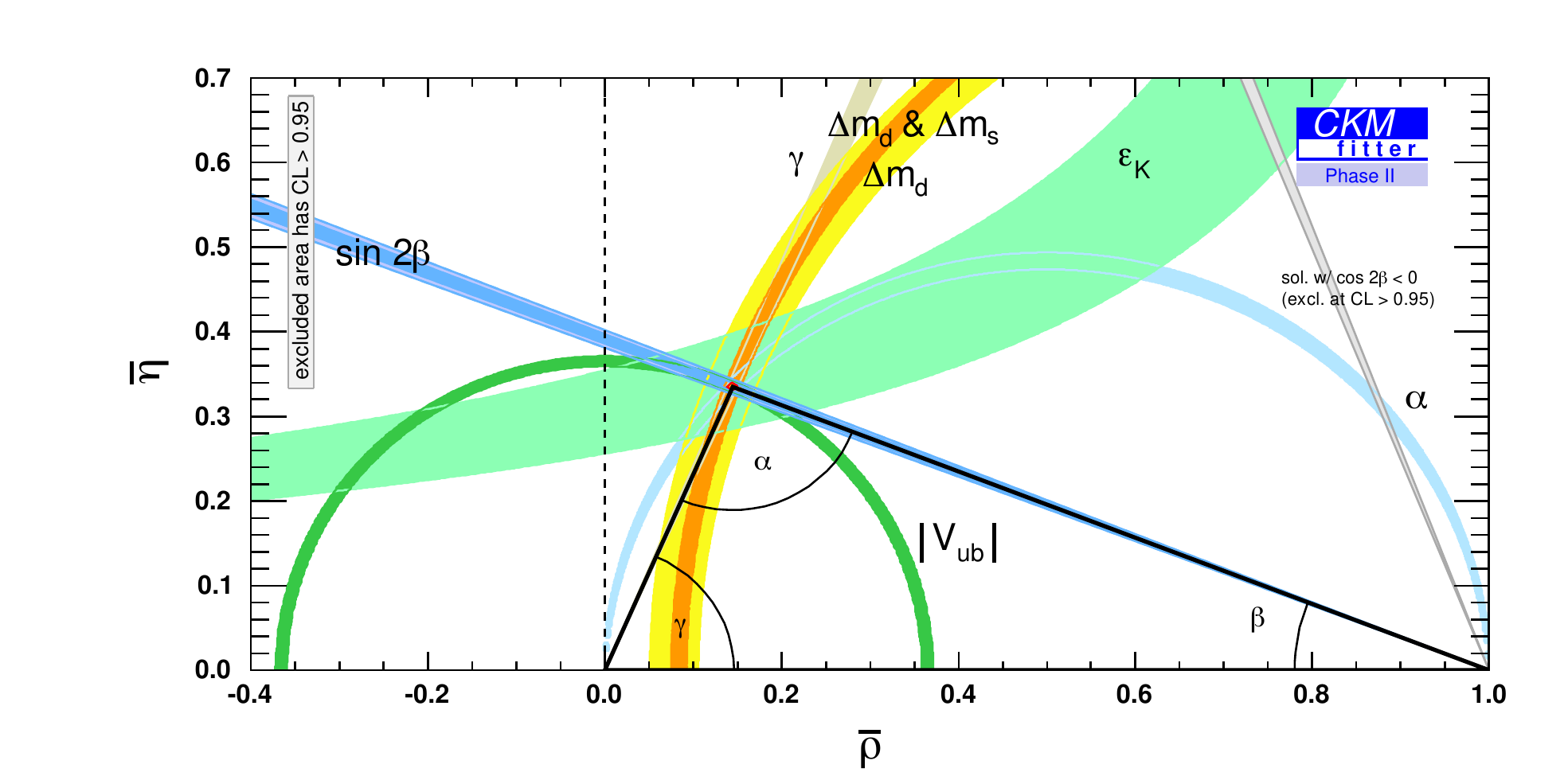}\\
\caption{The triangle, in the complex plane, representing the orthogonality of the first and third columns of the CKM matrix, \ie, $1+\frac{V_{tb}^*V_{td}}{V_{cb}^*V_{cd}}+\frac{V_{ub}^*V_{ud}}{V_{cb}^*V_{cd}}=0$, with (left) current and (right) future constraints overlaid~\cite{CKMfitter2015,Cerri:2018ypt}. The future constraints assume central values corresponding to a perfect agreement under the SM hypothesis and are based on expected sensitivities at Belle II (50\invab), BESIII, LHCb Upgrade II (300\invfb), and ATLAS/CMS (3\invab), and on improved theory inputs.\label{fig:CKMunitarity}}
\end{figure}

Precise tests of CKM unitarity remain crucial in constraining dynamics beyond the SM, particularly if new sources of \CP violation are present. Figure~\ref{fig:CKMunitarity} shows the current constraints on the determination of the apex of the standard CKM unitarity triangle in comparison with  the future constraints from expected measurements at Belle II, BESIII, LHCb Upgrade II and ATLAS/CMS, combined with future theory improvements \cite{CKMfitter2015,Cerri:2018ypt}.

The angle $\gamma$ is the only \CP-violation parameter of the SM that can be measured exclusively from tree-level processes, such as from the interference between $B^-\to\Dz(\to f) K^-$ and $B^-\to\Dzb(\to f) K^-$ decay amplitudes where $f$ is any final state directly accessible to both \Dz and \Dzb mesons. The theoretical uncertainties enter only at the level of one-loop electroweak corrections and are below $\mathcal{O}(10^{-6})$, because all the required hadronic matrix elements can be experimentally measured when enough final states $f$ are taken into account. New \CP violating effects in non-leptonic tree-level decays can modify the SM relation between $\gamma$ and the CKM elements by several degrees, making precise ($\lesssim1\degrees$) measurements of $\gamma$ powerful probes of NP~\cite{Brod:2014bfa,Lenz:2019lvd}. The present experimental uncertainty on $\gamma$ is about 4\degrees~\cite{HFLAV}, dominated by LHCb measurements~\cite{LHCb:2021dcr} with important inputs from CLEO and BESIII~\cite{CLEO:2010iul,BESIII:2020khq}, and it is limited by statistics. The experimental progress in the next decades should bring the uncertainty down to $0.35\degrees$, provided that strong-phase differences between \Dz and \Dzb amplitudes are measured with sufficient precision. The sample of coherent \mbox{$\Dz\!\Dzb$} pairs expected to be collected at BESIII will contribute $0.4\degrees$ to the $\gamma$ uncertainty. Hence, either larger samples of $\psi(3770)\to\Dz\!\Dzb$ data will need to be collected (\eg, at the STCF), or new methods to constrain these hadronic parameters will need to be developed.

\begin{figure}
\centering
\includegraphics[height=0.33\textheight]{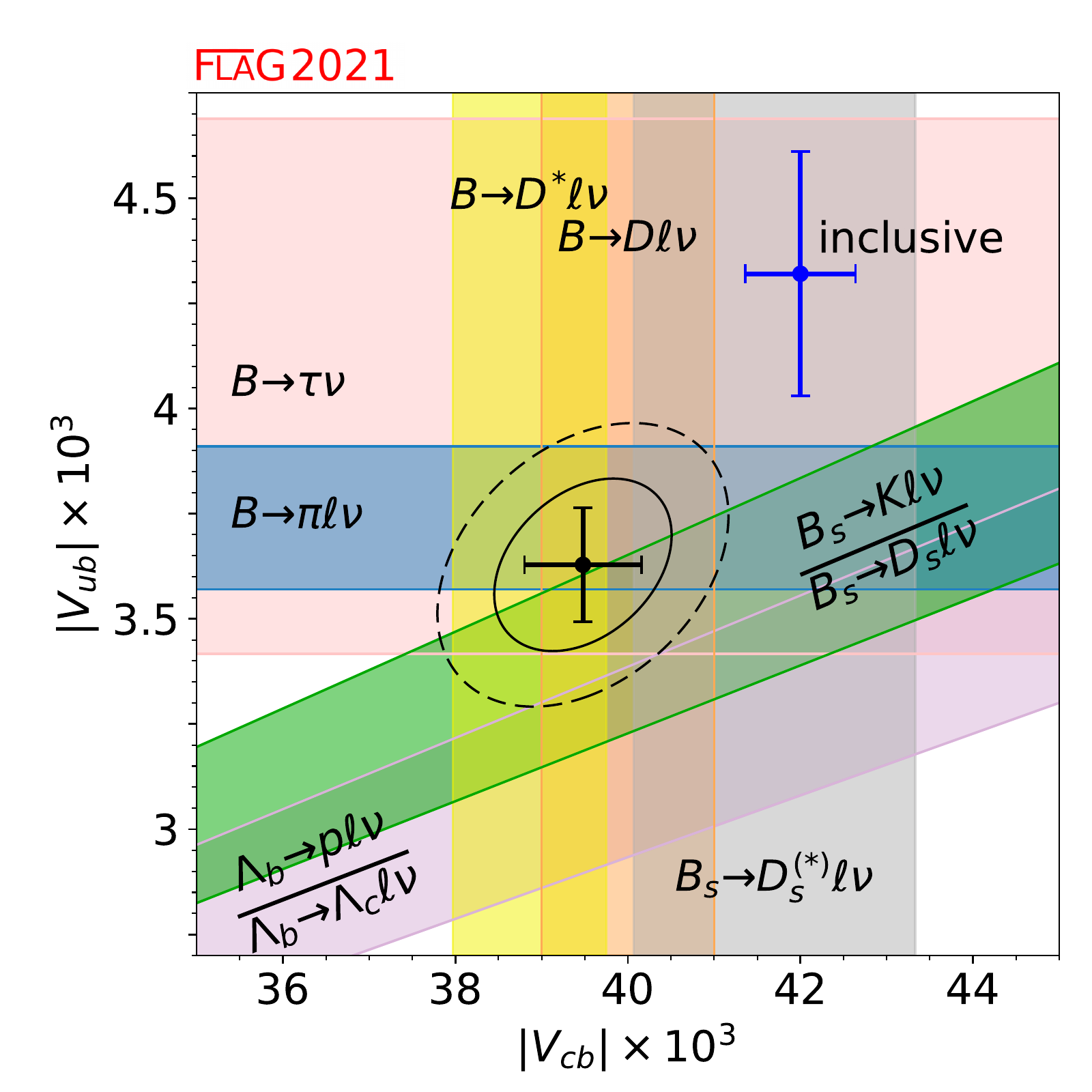}\hfil
\includegraphics[height=0.33\textheight]{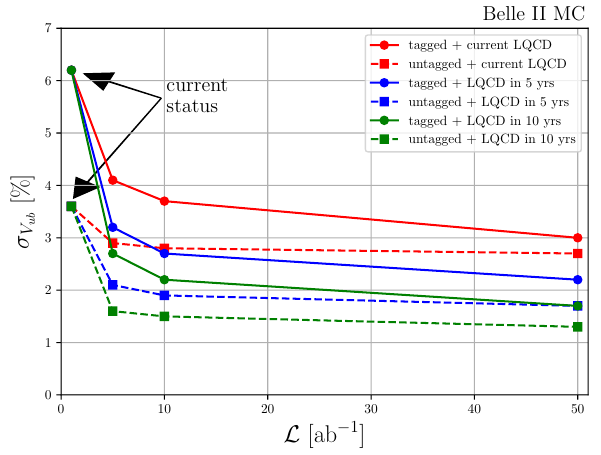}\\
\caption{(Left) Constraints in the $|V_{cb}|$-$|V_{ub}|$ plane from exclusive measurements and lattice QCD (colored bands, and their combination yielding the black point in the center) as well as inclusive measurements and operator-product expansions (blue point in the upper right)~\cite{Aoki:2021kgd}. (Right) Projections of uncertainties in exclusive measurements of $|V_{ub}|$ from  $\Bz\to\pi^-\ell^+\nu$ decays as functions of integrated luminosity at Belle~II~\cite{Belle-II:2018jsg}. Projections are separately made for analysis in which the partner $B$ meson is reconstructed (tagged) or not (untagged), and for current or future (expected) lattice QCD inputs\label{fig:VubVcb}\label{fig:Belle2-Bpilnu}.}
\end{figure}

The smallest and least well-known CKM matrix element magnitude is $|V_{ub}|$. In the standard unitarity triangle, $|V_{ub}|$ constrains the length of the side opposite to the precisely measured angle $\beta$ (Fig.~\ref{fig:CKMunitarity}). On the other hand, $|V_{cb}|$ normalizes the triangle and indirectly enters in SM predictions for many processes in flavor physics. For example, in the $b$ sector, $\BF(\Bs\to \mu^+\mu^-)$ is approximately proportional to $|V_{cb}|^2$; in the kaon sector, $\BF(\KL\to\pi^0\nu\nub)$ and $\epsilon_K$ behave like $|V_{cb}|^{4}$ and $|V_{cb}|^{3.4}$, respectively~\cite{Buras:2021nns}. For both $|V_{ub}|$ and $|V_{cb}|$, there are persistent tensions between exclusive and inclusive determinations from semileptonic $b$-hadron decays, as shown in Fig.~\ref{fig:VubVcb}. The reason for this discrepancy is unknown. Most indications point to possibly inconsistent experimental or theory inputs, but interpretations in terms of non-SM physics cannot be excluded~\cite{Gambino:2020jvv}. Improving both the exclusive and inclusive determinations of $|V_{ub}|$ and $|V_{cb}|$ is a high priority that will require a combined experiment-theory effort. Experimentally, Belle II will drive the global progress throughout the next decades. With the 50\invab of integrated luminosity, it will achieve $\mathcal{O}(1\%)$ precision on inclusive $|V_{ub}|$ and will double the global precision in exclusive $|V_{ub}|$ results, independently of any improvement in theoretical inputs. With the same sample, Belle II will reach $\mathcal{O}(1\%)$ precision on both inclusive and exclusive determinations of $|V_{cb}|$. Expected progress in lattice QCD will then offer further significant improvement as, \eg, shown in Fig.~\ref{fig:Belle2-Bpilnu} for $|V_{ub}|$, and as discussed in Sec.~\ref{sec:theory-progress}. Unique contributions from LHCb will come from measurements of rates of a variety of $b$-hadron decays. LHCb has already performed measurements of $|V_{cb}|$ using $\Bs\to\DsorDssm\mu^+\nu$ decays and of the ratio $|V_{ub}|/|V_{cb}|$ using semileptonic \Lb and \Bs decays~\cite{LHCb:2020cyw,LHCb:2015eia,LHCb:2020ist}. These offer complementary sensitivity, particularly because they are subject to different theory inputs. The planned detector improvements in Upgrade II will significantly enhance the opportunities for more studies based on other \Bs decays modes and on $B^+_c$ mesons.

Determinations of $|V_{cs}|$ and $|V_{cd}|$ rely on rate measurements of leptonic and semileptonic charm decays performed by CLEO, BaBar, Belle, and BESIII, together with lattice calculations of the corresponding form factors or decay constants~\cite{HFLAV,USQCD:2019hyg,Aoki:2021kgd,Boyle:2022uba,USQCD:2022mmc}. The most precise determinations yield $1\%$-level uncertainties, with sub-percent precision recently achieved for $|V_{cs}|$ from $D \to \Kbar \ell^+\nu$ \cite{Chakraborty:2021qav}. The precision with semileptonic decays is limited by the lattice-QCD computation of the decay form-factors. Measurements with leptonic decays are, instead, limited by experimental uncertainties. The relative uncertainties in $|V_{cs}|$ and $|V_{cd}|$ from leptonic decays are expected to be reduced at BESIII from 2.6\% and 1.2\% to approximately 1.1\% and 0.9\%, respectively~\cite{BESIII-Whitepaper}. Further improvement will be possible when combining with measurements at Belle II~\cite{Belle-II:2018jsg} and at the STCF, provided that systematic uncertainties can be reduced well below the 1\% level~\cite{STCF-Whitepaper}.

\begin{figure}
\centering
\includegraphics[width=0.5\textwidth]{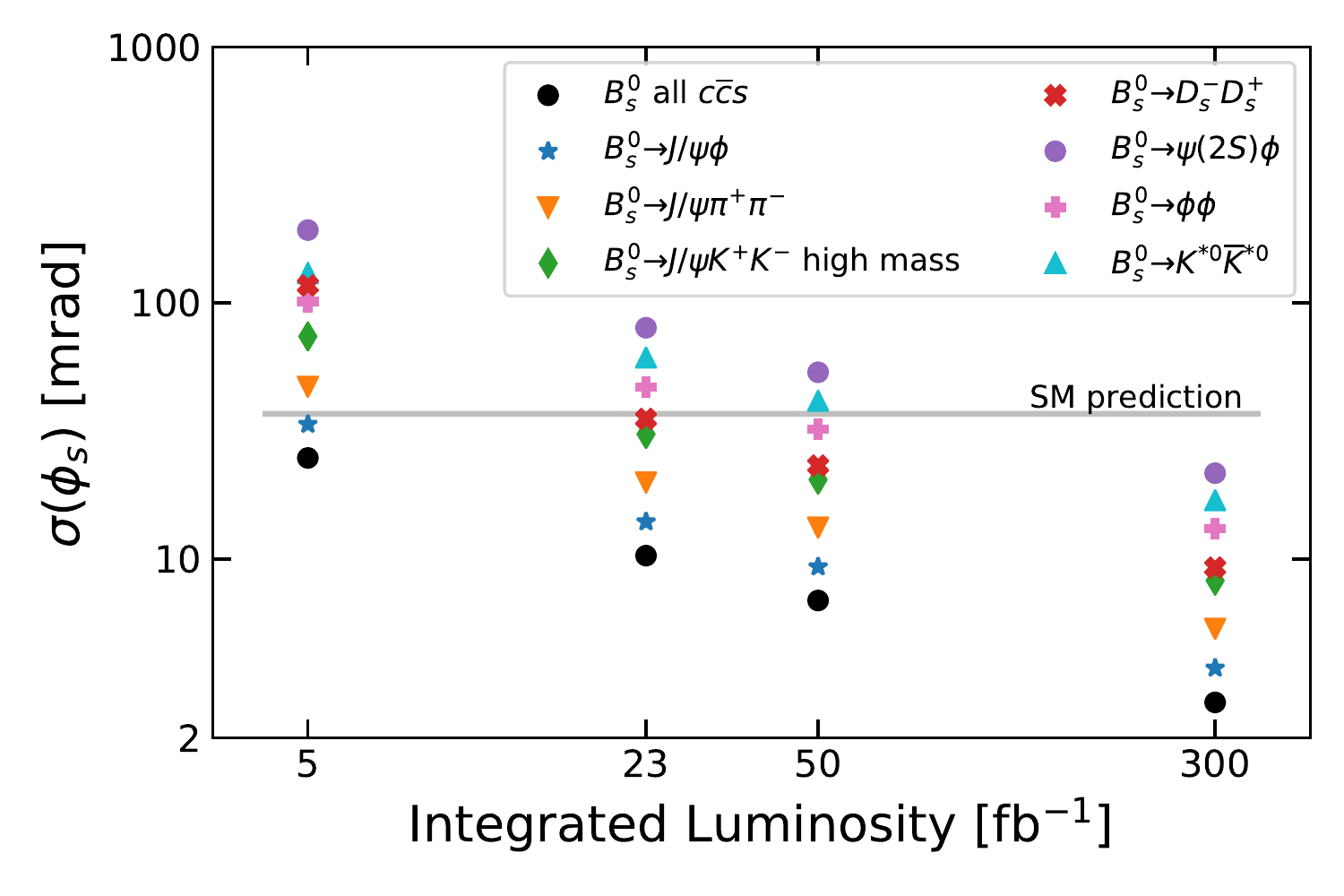}
\caption{Projected sensitivity at LHCb for $\phi_s$ using various decay modes~\cite{LHCb-TDR-023}. The SM prediction and its uncertainty~\cite{CKMfitter2015} is also shown as the grey band.\label{fig:LHCb-phis}}
\end{figure}

New sources of \CP violation departing from the CKM picture can be effectively searched for in neutral meson mixing, where the knowledge of the \CP-violating mixing phases are limited by the experimental statistical uncertainties. This is particularly true for the \Bs-\Bsb mixing phase $\phi_s\approx-2\beta_s$, which is suppressed by a factor $0.05$ compared to the \Bz-\Bzb phase $\beta$. The world average value, based on measurements from the LHC and Tevatron experiments, has a precision of 19\mrad~\cite{HFLAV}. The SM prediction, taken as the indirect determination of $-2\beta_s$ via the global CKM fit to experimental data, has an uncertainty of 0.8\mrad~\cite{CKMfitter2015}. Such estimation, however, neglects contribution to $\phi_s$ from subleading penguin amplitudes that are estimated to be smaller than 21\mrad~\cite{Frings:2015eva}. Improved measurements of $\phi_s$ will be performed at LHCb, ATLAS and CMS, with a combined projected uncertainty based exclusively on the $\Bs\to\jpsi\phi$ channel estimated to go below 4\mrad. To possibly expose NP, the expected progress in experimental precision will require improved constraints on the SM penguin contributions, which can be achieved by measuring $SU(3)$-related decays~\cite{LHCb:2014xpr,LHCb:2015esn,LHCb:2018roe}. Other avenues for beyond-SM sources of \CP violation are penguin-dominated $b\to q\bar{q}s$ decays such as $\Bs\to\phi\phi$ or $\Bs\to K^{*0}\Kbar{}^{*0}$. The sensitivities expected at LHCb on these channels are shown in Fig.~\ref{fig:LHCb-phis}. Similarly, in the \Bz system, time-dependent \CP asymmetries in $\Bz\to\phi\KS$ and $\Bz\to\eta'\KS$ can be compared to determinations of $\sin2\beta$ based on tree-dominated $b\to c\bar{c}s$ transitions, such as $\Bz\to\jpsi\KS$~\cite{CPV-Overview-Whitepaper}.

\begin{figure}[t]
\centering
\includegraphics[width=\textwidth]{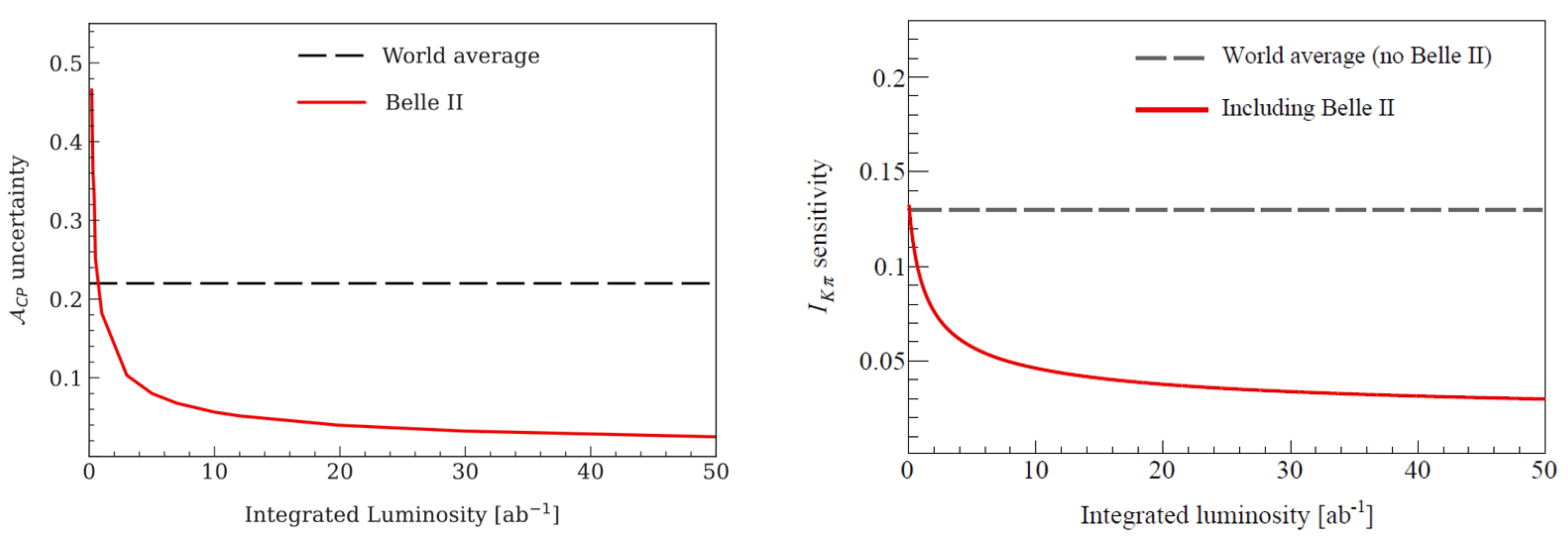}\\
\caption{Projected uncertainty as a function of the expected Belle~II sample size in (left) the decay-time-integrated \CP asymmetry of $B^0\to\pi^0\pi^0$ decays and (right panel) in the $B\to K\pi$ isospin sum rule $I_{K\pi}$. The solid red curve shows the projection assuming updates on the complete set of $B\to K\pi$ measurements. The dashed grey curve represents the projection assuming no Belle II inputs~\cite{BelleII-Whitepaper}.\label{fig:BelleII-charmless}}
\end{figure}

Charmless $B$ decays give access to $\alpha$, the least known angle of the CKM unitarity triangle, which also suffers from much larger theory uncertainties compared to other CKM angles. Appropriate combinations of measurements from decays related by isospin symmetries, such as $\Bz\to(\pi\pi)^0$, $(\rho\pi)^0$, $(\rho\rho)^0$ and $a_1^\pm\pi^\mp$, reduce the impact of hadronic uncertainties and yield a robust direct determinations of $\alpha$ with a 4\degrees uncertainty~\cite{HFLAV}. The determination of $\alpha$ is expected to be dominated by Belle II with a projected uncertainty of $0.6\degrees$ with 50\invab, provided that there is also an improved understanding of the size of isospin breaking (\eg, using $B\to\pi\eta^{(')}$ decays). Isospin symmetry also provides so-called sum rules, which are linear combinations of branching fraction and \CP asymmetries that offer null tests of the SM. An interesting case is provided by the so-called $K\pi$ puzzle, a long-standing $3\sigma$-deviation anomaly associated with the difference between direct \CP asymmetries in $\Bz\to K^+\pi^-$ and $B^+\to K^+\pi^0$ decays, and more generally from the isospin sum rule $I_{K\pi}$ relating $\Bz\to K^+\pi^-$, $B^+\to K^0\pi^+$, $B^+\to K^+\pi^0$ and $\Bz\to K^0\pi^0$ decays~\cite{Gronau:2005kz}. Since $I_{K\pi}$ is predicted to be zero within $\mathcal{O}(1\%)$ in the SM, a precise determination of all inputs offers a reliable and precise null test of the SM. The current experimental sensitivity of $\sim13\%$, limited by the BaBar and Belle measurements of the $B^0\to K^0\pi^0$ mode, is expected to be improved by Belle II in the next decade as shown in Fig.~\ref{fig:BelleII-charmless}.

Unique opportunities to search for new sources of \CP violation in the up-type-quark sector are provided by the study of charm hadrons. LHCb made the first observation of \CP violation in charm decays in 2019 by measuring a nonzero difference in the \CP asymmetries of $D^0\to K^+K^-$ and $D^0\to\pi^+\pi^-$ decays, $\Delta\ACP$~\cite{LHCb:2019hro}. In contrast to $b$ decays, loop amplitudes in charm are severely suppressed by the GIM mechanism and SM \CP violation arises mostly from the interference of tree-level amplitudes, possibly associated with rescattering~\cite{Grossman:2019xcj,Bediaga:2022sxw}. Rescattering amplitudes are challenging to compute and make the interpretation of the observed \CP violation ambiguous. Precise measurements of \CP asymmetries in other decay channels help constrain the rescattering effect and are therefore crucial to understand the underlying dynamics~\cite{CPV-Overview-Whitepaper}. The Cabibbo-suppressed decay $D^+\to\pi^+\pi^0$ is a particularly interesting mode as it proceeds essentially via a single tree amplitude in the SM~\cite{Buccella:1992sg,Grossman:2012eb}. Thus, observing direct \CP violation in $D^+\to\pi^+\pi^0$ would be a robust indication of new dynamics. While LHCb will continue to dominate the precision on \CP asymmetries in decay modes with charged particles in the final state~\cite{LHCb:2018roe}, in the next decade Belle~II is expected to dominate the precision on $\ACP(D^+\to\pi^+\pi^0)$ and will be the only experiment able to precisely measure modes with only neutrals~\cite{BelleII-Whitepaper}, such as $\Dz\to\pi^0\pi^0$ in which \CP violation is expected to be of a measureable size~\cite{Wang:2022nbm}.

\begin{figure}
\centering
\includegraphics[width=0.6\textwidth]{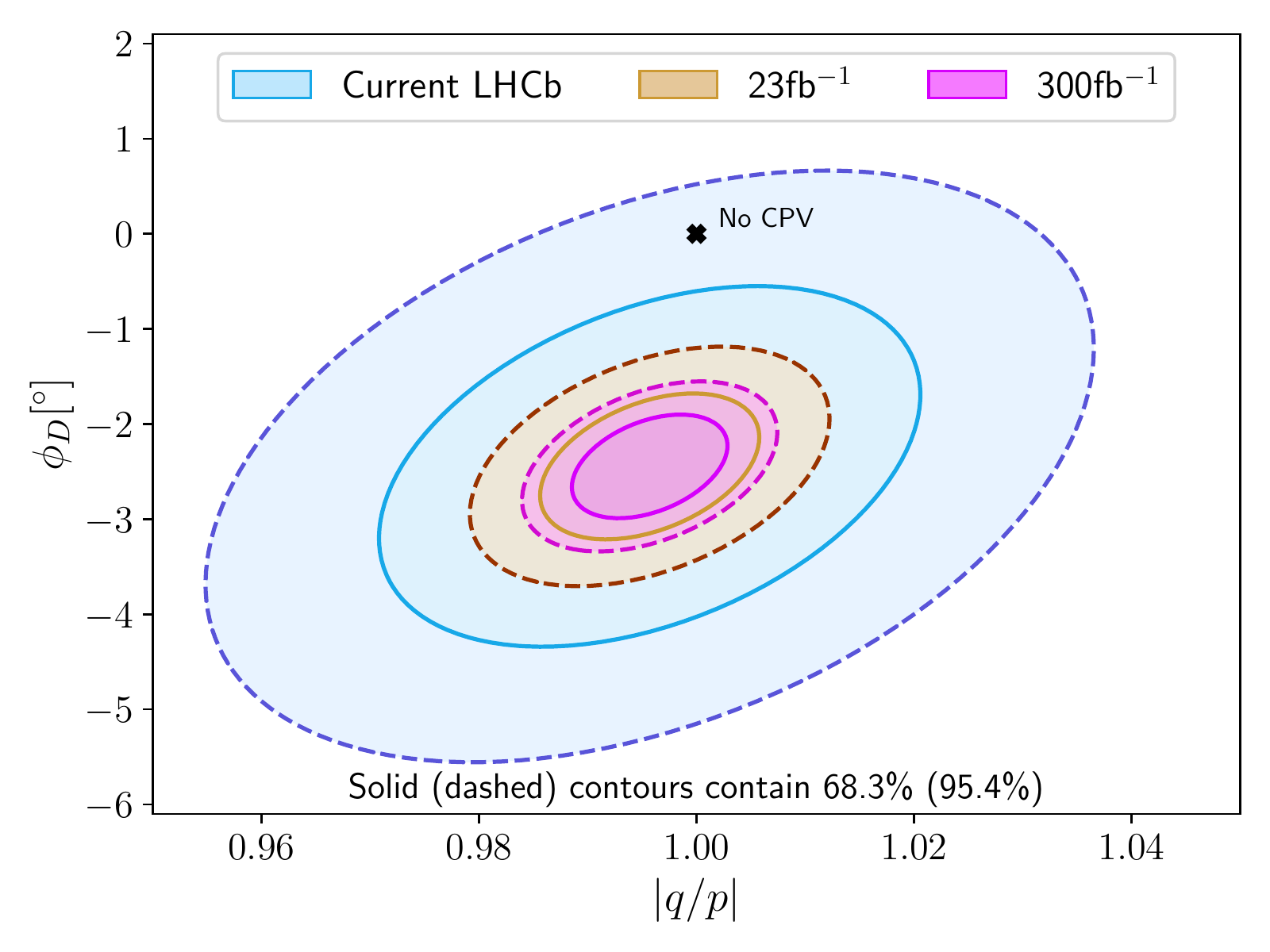}
\caption{Projected sensitivity at LHCb to the parameters of \CP violation in charm mixing, $|q/p|$ and $\phi_D$, assuming the current central values of experimental observables~\cite{LHCb-TDR-023}. Contours shaded with different levels of darkness indicate 68.3\% and 95.4\% confidence-level regions.\label{fig:LHCb-charm}}
\end{figure}

LHCb's recent observation of a nonzero mass difference between neutral charm eigenstates~\cite{LHCb:2021ykz} paves the the way for future precision measurements of mixing parameters and searches of \CP-violation effects in \Dz-\Dzb mixing. The current constraints on \CP violation in charm mixing are at least an order of magnitude above the expected SM contribution and are limited by statistical experimental uncertainties~\cite{Bobrowski:2010xg,Kagan:2020vri,HFLAV}. The data sets expected from the operation of the Upgrade I and Upgrade II detectors make LHCb the only planned experiment with a realistic possibility of observing \CP violation in charm mixing, since it has the best sensitivity to decay modes such as $\Dz\to K^-\pi^+$, $\Dz\to K^+K^-$, $\Dz\to\pi^+\pi^-$ and $\Dz\to\KS\pi^+\pi^-$~\cite{LHCb:2018roe}. As examples, projected sensitivities for the mixing-induced \CP-violation observables $A_\Gamma$ in $\Dz\to K^+K^-,\pi^+\pi^-$ and $\Delta x$ in $\Dz\to\KS\pi^+\pi^-$ are reported in Tab.~\ref{tab:sensitivity}; the projected sensitivity for the parameters $|q/p|$ and $\phi_D$, resulting from the combination of these and other measurements at LHCb, is shown in Fig.~\ref{fig:LHCb-charm}.

\section{Farther into the future\label{sec:far-future}}
The emergence of high-energy $e^+e^-$ circular collider projects to accurately study the properties of the Higgs boson opens a new appealing perspective for a continued heavy-quark-physics program past the HL-LHC era~\cite{Bhat:2022hdi,Liu:2022rua,FCC-Whitepaper,CEPC-Whitepaper,EF-Report}. The proposed machines will operate at all the relevant electroweak thresholds ($Z^0$, $H^0$, $W^+W^-$, $t\bar{t}$) and will give access to abundant samples of $b$- and $c$-hadron decays. As an example, the expected number of $Z^0$ decays to be collected with FCC-ee will provide about 20 times more $B$ mesons, and about 9 times more charm hadrons, than expected at Belle II with 50\invab~\cite{FCC-Whitepaper}. Moreover, similarly to LHCb, all $b$-flavored particles will be produced and with a significant boost to allow precise measurements of decay-time-dependent observables.

FCC-ee and/or CEPC will be able to complement many of the studies performed at Belle II and LHCb, and significantly extend the program in several critical areas. A particular strength will be the ability to make sensitive studies of modes containing neutral hadrons, photons and neutrinos, with much larger sample sizes than will be available at Belle II. This possibility will enable FCC-ee/CEPC to harness a wide range of charm-meson decay modes in measurements of $\gamma$ from $B^-\to DK^-$ and $\Bs\to\DsorDssm K^+$ decays. It is expected that the flavour-tagging efficiency will be significantly higher than at the LHC, bringing corresponding gains for time-dependent measurements of \Bs decays~\cite{Aleksan:2021gii,Li:2022tlo}. Other interesting possibilities include modes relevant for the angle $\alpha$; \eg, precise measurements of the time-dependent \CP asymmetries in $\Bz\to\pi^0\pi^0$ can be performed making use of both the $\gamma\gamma$ and $e^+e^-\gamma$ decays of the $\pi^0$ meson. At a high-energy $e^+e^-$ collider, another approach will open up for precise measurements of CKM matrix elements that has no systematic limitation due to the knowledge of hadronic inputs, \eg from lattice QCD: direct determination from hadronic decays of $W^+$ bosons. Several $10^8$ $W^+$ boson decays will be collected when operating FCC-ee and/or CEPC at the $W^+W^-$ threshold and above, making possible, \eg, measurements of $|V_{cb}|$ with up to an order of magnitude improved precision with respect to present results. In addition to the measurement of CKM-related observables, FCC-ee/CEPC will perform studies of a wide range of suppressed flavor-changing neutral-current processes, such as $b\to s(d)\ell^+\ell^-$, $b\to s(d)\tau^+\tau^-$ and $b\to s(d)\nu\nub$ (see, \eg, Refs.~\cite{Kamenik:2017ghi,Li:2020bvr,Li:2022tov}). This program will be extended to the analysis of favored, but experimentally challenging, modes, where the SM predictions are reliable and beyond-SM effects could be pronounced (\eg, the decays $B_c^+\to\mu^+\nu$ and $B_c^+\to\tau^+\nu$~\cite{Zheng:2020ult,Amhis:2021cfy}). The analysis of these channels, together with that of radiative flavor-changing neutral currents in both the beauty and charm sectors, will provide stringent tests of the SM and have high discovery potential for NP. Moreover, baryons with $b$ and $c$ quarks from $Z^0$ decays are strongly longitudinally polarized, giving access to new types of angular observables~\cite{Hiller:2001zj}.

Measurements of weak decays of $b$ and $c$ quarks place specific demands on the detector design, particularly in the areas of vertexing, calorimetry and particle identification. These requirements are not necessarily the same as those required for electroweak and Higgs physics, and motivate a machine with four interaction points with one experiment devoted to heavy-flavor physics.

With a Higgs factory likely to represent the medium-term future of particle physics, the long-term future of the field crucially depends on the next generation of high-energy colliders to push forward the reach for direct production of NP particles~\cite{EF-Report}. Precise measurements of flavor observables in the next decades are likely to provide unique inputs that can have a major impact on the motivation and planning of such facilities. As an example, NP in $b\to s\mu^+\mu^-$ transitions would suggest that a muon collider of 10\tev (or higher) has greater potential for a direct discovery than a 100\tev $pp$ collider~\cite{Altmannshofer:2022xri,Azatov:2022itm}.

\section{Expected theory progress\label{sec:theory-progress}}
The expected experimental advances in weak decays of $b$ and $c$ hadrons must be complemented by commensurate advances in theoretical calculations. In many cases, measurements involving hadrons cannot be related to the underlying short-distance physics of interest (such as CKM matrix elements) without calculations of hadronic matrix elements, using lattice QCD or other approaches, such as the operator product expansion (OPE) for inclusive observables. In addition, the Wilson coefficients of weak effective operators need to be calculated in the SM and in NP models, and efforts in model building are required to find ultraviolet-complete explanations of observed deviations from the SM that ideally also address some of the other fundamental questions about the Universe outside of flavor physics.

While for some observables, such as purely leptonic $\Bz_{(s)}$-meson decay rates, theoretical predictions are currently more precise than experiment, the opposite is true for other observables, such as the \CP asymmetries in hadronic charm decays where reliable SM predictions are still unavailable. For the determinations of $|V_{ub}|$ and $|V_{cb}|$ in semileptonic $b$-hadron decays, the uncertainties are currently more evenly split between experiment and theory, and both need to improve at a similar pace. 

The following subsections summarize the status and prospects of theoretical calculations for selected important quantities or processes. Many of these topics are also discussed in the Theory-Frontier topical-group-6 report \cite{TF06-Report}.

\subsection{Exclusive leptonic and semileptonic decays\label{sec:theory:exclusivesemilept}}
At leading order in weak effective theory and at leading order in QED, the nonperturbative QCD contributions to exclusive leptonic or semileptonic decays are given by matrix elements of the form $\langle 0|\bar{q}^\prime \Gamma q |h\rangle$ or $\langle h^\prime|\bar{q}^\prime \Gamma q |h\rangle$, respectively, where $h$ and $h^\prime$ denote the hadrons in the initial and final states. These matrix elements can be expressed in terms of Lorentz-scalar decay constants or form factors, respectively, and are required inputs in the determination of $|V_{ub}|$, $|V_{cb}|$, $|V_{cd}|$, $|V_{cs}|$, in the theory predictions for lepton-flavor universality ratios, and in the theory of rare $b$ and $c$ decays. In general, studying the quark-level transitions of interest in multiple different decay channels involving hadrons with different spin helps in disentangling different operator structures of NP couplings. For example, the baryonic decay $\Lb\to p \mu^-\nub$ provides constraints on right-handed couplings in the $b\to u$ weak effective Hamiltonian which are complementary to those from $\Bbar\to\pi\mu^-\nub$~\cite{LHCb:2015eia,Detmold:2015aaa,Albrecht:2017odf}.
Furthermore, theoretical predictions may also be needed for transitions to higher-lying states that contribute to the backgrounds in measurements, such as $B\to D^{**} \ell^+\nu$ \cite{Bernlochner:2016bci,Bernlochner:2017jxt,Bernlochner:2021vlv}.

Numerical lattice-gauge-theory computations now provide the most precise results for decay constants and for many form factors, and the precision can be improved even more in the future \cite{USQCD:2019hyg,Aoki:2021kgd,Boyle:2022uba,USQCD:2022mmc}. Light-cone sum rules and QCD factorization/soft-collinear effective theory (see, \eg, Refs.~\cite{Ball:2004rg,Wang:2015vgv,Wang:2015ndk,Cheng:2017smj,Gubernari:2018wyi,Feldmann:2018kqr,Descotes-Genon:2019bud,Bordone:2019vic,Gao:2021sav,Khodjamirian:2022vta}) can be used for final states or kinematic regions that are challenging for lattice QCD. In addition, heavy-quark effective theory (HQET) can provide useful relations, especially for $b\to c$ transitions (see, \eg, Refs.~\cite{Bordone:2019vic,Bernlochner:2018bfn}); here it is anticipated that long-standing questions about higher-order corrections can be answered with new experimental and lattice results in the coming years \cite{Gambino:2020jvv}. Another important theoretical tool is dispersion relations based on analyticity and unitarity, which yield bounds on form factors that can stabilize kinematic extrapolations \cite{Caprini:2019osi}.

For the charm-meson decay constants $f_D$ and $f_{D_s}$, the current averages of lattice results in pure QCD and in the isospin limit have total uncertainties of 0.3\% and 0.2\%, respectively, while the uncertainties for $f_B$ and $f_{B_s}$ are 0.7\% and 0.6\%, respectively. Key in achieving this precision was the use of the same type of relativistic lattice action for the heavy quarks as used for the light quarks, which eliminates the systematic uncertainty associated with the matching of the vector and axial-vector currents from the lattice to the continuum. For bottom quarks, this approach still requires extrapolations in the heavy-quark mass, as most lattices used in the analyses have a spacing $a$ that is not small enough to satisfy $a m_b<1$. For semileptonic form factors, first calculations using the fully relativistic approach have been performed for $D\to K$ \cite{Chakraborty:2021qav,Parrott:2022rgu}, $B\to K$ \cite{Parrott:2022rgu}, $B\to\DorDs$ \cite{Kaneko:2021tlw}, $\Bs\to\DsorDss$ \cite{McLean:2019qcx,Harrison:2021tol}, $B^+_c\to\jpsi$ \cite{Harrison:2020gvo}, and $B\to\pi$ \cite{Colquhoun:2022atw}. Reaching sub-percent precision for the $b$-hadron semileptonic form factors, will require large numbers of samples (gauge configurations/source locations) to reduce the statistical uncertainties (especially for $B\to\pi$, which is intrinsically noisier than, \eg, $\Bs\to D_s^-$) and ultrafine lattices, demanding leadership-class computing resources~\cite{Boyle:2022ncb}.

If the hadron in the final state is a resonance with non-negligible decay width, a rigorous theoretical treatment requires computing transition matrix elements to the multi-hadron asymptotic final states to which the resonance couples, followed (if desired) by analytic continuation to the resonance pole. The mathematical formalism (known as the Lellouch-L\"uscher formalism) relating the infinite-volume transition matrix elements of interest with the finite-volume transition matrix elements accessible on the lattice is well-established for $1\to 2$ transitions \cite{Luscher:1986pf,Luscher:1991cf,Lellouch:2000pv,Lin:2001ek,Christ:2005gi,Hansen:2012tf,Briceno:2014uqa,Briceno:2015csa,Agadjanov:2016fbd,Briceno:2021xlc} and has already been implemented to perform lattice calculations of $K\to\pi\pi$ weak decays \cite{Blum:2011ng,RBC:2015gro,Ishizuka:2018qbn,RBC:2020kdj,RBC:2021acc} and of the electromagnetic $\pi\gamma^\ast\to \rho (\to \pi\pi)$ transition \cite{Briceno:2015dca,Briceno:2016kkp,Alexandrou:2018jbt}. The formalism is also applicable to semileptonic decays with two-body resonances in the final state, such as $B\to\rho(\to\pi\pi)\ell^+\nu$, $B\to K^*(892)(\to K\pi)\ell^+\ell^-$, $B\to K_0^*(700)(\to K\pi)\ell^+\ell^-$. Here, it is possible to compute the $B\to \pi\pi$ or $B\to K\pi$ transition form factors for the relevant partial waves (\eg, $P$ wave and $S$ wave) as functions of both the dilepton invariant mass squared, $q^2$, and the $\pi\pi$ or $K\pi$ center-of-momentum energy squared, $s$. The analysis must take into account rescattering into other multi-hadron channels if $s$ is above their thresholds, and neglecting such channels thus places an upper limit on the accessible range of $s$. It is known how to take into account multiple coupled two-body channels (\eg, $\pi\pi+K\Kbar$), and work is underway to extend the formalism to the three-body and higher sectors \cite{Muller:2020wjo,Hansen:2021ofl}. Dispersion theory can also be used to describe the $s$ dependence, see \eg~\cite{Kang:2013jaa}.

\subsection{QED corrections to leptonic and semileptonic decays\label{sec:theory:QEDleptonic}}
With QCD uncertainties reduced to the sub-percent level, further theoretical improvements are needed in the treatment of QED corrections. This applies in particular to soft or collinear photons that can lead to large logarithms, and to hadron-structure-dependent effects, neither of which are accounted for by the commonly included Sirlin factor $\eta_{\rm EW}$ \cite{Sirlin:1981ie}. Soft photon radiation in experiments is modelled using PHOTOS~\cite{Golonka:2005pn}, which however neglects radiation from charged initial-state particles and other important effects \cite{Szafron-CKM}.

Significant progress in the treatment of QED corrections has been made recently in effective-field-theory-based approaches and factorization \cite{Beneke:2017vpq,Beneke:2019slt,Isidori:2020acz,Beneke:2021jhp,Isidori:2022bzw,Szafron-CKM,Zwicky:2021olr}. Sizeable hard-collinear logarithms were identified and were shown to be absent at the structure-dependent level using gauge invariance \cite{Isidori:2020acz}. Factorization in QCD$\times$QED is still a rather new subject under active development. It must be noted that, as soon as non-perturbative soft matrix elements are evaluated in factorization including QED, light-cone distribution amplitudes need to be generalized accordingly \cite{Beneke:2021pkl}.

First lattice-QCD calculations of structure-dependent QED corrections to leptonic decays have been performed for pion and kaon leptonic decays \cite{Giusti:2017dwk,DiCarlo:2019thl}, and this approach is in principle also applicable to $B_{(s)}$ and $D_{(s)}$ leptonic decays. In Refs.~\cite{Giusti:2017dwk,DiCarlo:2019thl}, real-photon emission was treated in the point-like approximation, which is sufficient for $\pi$ and $K$ decays to muons, but not for decays to electrons. Since then, full structure-dependent lattice calculations of real-photon emission in leptonic decays have also been performed \cite{Kane:2019jtj,Desiderio:2020oej,Frezzotti:2020bfa,Kane:2021zee}. These calculations are interesting not only in the context of QED corrections to leptonic decays but can also describe radiative leptonic decays with hard photons. The hard photon lifts the helicity suppression and, in the case of $B^0_{(s)}\to\ell^+\ell^-\gamma$, provides sensitivity to a larger set of operators in the weak effective Hamiltonian. Furthermore, radiative leptonic $B$ decays at high photon energy are well suited to constrain the first inverse moment of the $B$-meson light-cone distribution amplitude, an important parameter in the theory of nonleptonic $B$ decays \cite{Korchemsky:1999qb,Beneke:1999br,DescotesGenon:2002mw,Lunghi:2002ju,Braun:2012kp,Wang:2016qii,Beneke:2018wjp,Wang:2018wfj,Shen:2018abs,Khodjamirian:2020hob,Beneke:2020fot,Shen:2020hsp,Carvunis:2021jga,Lu:2021ttf,Janowski:2021yvz,Pullin:2021ebn}.

Recently, lattice calculations have also been demonstrated for $P\to\ell\nu\ell^{\prime+}\ell^{\prime-}$, where the $\ell^{\prime+}\ell^{\prime-}$ pair is produced through a virtual photon \cite{Tuo:2021ewr,Gagliardi:2022szw}. Work is also underway to compute structure-dependent QED corrections to semileptonic decays on the lattice, which is substantially more challenging compared to leptonic decays \cite{Sachrajda:2019uhh}. The electroweak box diagrams have also been computed on the lattice for kaon semileptonic decays \cite{Ma:2021azh}. Further progress with QED corrections to semileptonic decays, and extensions of the calculations from light/strange mesons to charm and bottom mesons, would be desirable.

\subsection{Exclusive rare $b$ and $c$ hadron decays}\label{sec:raredecaystheory}
The theory uncertainties for rare $b$ and $c$ decays vary widely among different types of processes and observables \cite{Rare-decay-Overview-Whitepaper,LFUV-Overview-Whitepaper}. Muon-to-electron lepton-flavor-universality ratios such as $R_K$ are predicted to be very close to unity in the SM, and one of the main sources of uncertainty is QED, as discussed in the previous Section and in Ref.~\cite{LFUV-Overview-Whitepaper}. The branching fractions of the purely leptonic $B^0_{(s)}\to\mu^+\mu^-$ are also predicted quite precisely, thanks to sub-percent-precision lattice-QCD results for the decay constants \cite{Bazavov:2017lyh} and recent progress with perturbative QCD and QED corrections \cite{Beneke:2019slt}; the dominant source of uncertainty in these branching fractions is now $|V_{cb}|$.

The theory of semileptonic $b\to s\ell^+\ell^-$ branching fractions and angular observables depends on both local hadronic matrix elements of quark-bilinear currents, described by local form factors, and nonlocal hadronic matrix elements involving four-quark or quark-gluon operators together with the quark electromagnetic current at a different spacetime point. Both are important sources of uncertainty. Higher-precision lattice calculations of the local form factors are expected in the future as discussed in Sec.~\ref{sec:theory:exclusivesemilept}. For the nonlocal matrix elements, the charm contributions are the most significant and problematic. At high $q^2$, the nonlocal matrix elements may be approximated using a local OPE \cite{Grinstein:2004vb,Beylich:2011aq}, but the OPE is unable to predict the detailed $q^2$-dependence associated with broad charmonium resonances in this region. At low $q^2$, the available approaches include QCD factorization \cite{Beneke:2001at} and the light-cone OPE \cite{Khodjamirian:2010vf,Khodjamirian:2012rm}. The latter calculation was recently further improved and combined with dispersive bounds \cite{Gubernari:2020eft,Gubernari:2022hxn}. The theory of the nonlocal matrix elements at low $q^2$ is more challenging for \Lb decays, where new types of nonfactorizable  spectator-scattering contributions arise \cite{Wang:2014jya,Feldmann:2021zta}. In contrast to semileptonic modes with charged leptons, dineutrino modes ($b\to s\nu\nub)$ do not receive contributions from these nonlocal matrix elements and can be predicted precisely using just the local form factors. In $b\to d\ell^+\ell^-$ decays, non-local effects are qualitatively different as contributions from $\rho$ and $\omega$ resonances are not suppressed~\cite{Rare-decay-Overview-Whitepaper,Hambrock:2015wka,Khodjamirian:2017fxg}. Controlling those effects will be crucial to establish exclusive $b \to d \ell^+\ell^-$ decays as important probes of NP~\cite{Rusov:2019ixr}.

Rare charm decays involving the transitions $c\to u \ell^+\ell^-$ or $c\to u\gamma$ are subject to large long-distance contributions that dominate the short-distance contributions by orders of magnitude \cite{Burdman:2001tf,Fajfer:2015mia,deBoer:2015boa,Bharucha:2020eup}. Nevertheless, as already mentioned, one can construct observables that are strongly suppressed in the SM and serve as clean null tests, such as \CP asymmetries and angular observables that vanish in the SM~\cite{deBoer:2017que,Meinel:2017ggx,DeBoer:2018pdx,Bause:2019vpr,Golz:2021imq,Golz:2022alh}. Additional probes for NP are lepton-flavor-violating transitions and dineutrino modes \cite{Bause:2019vpr,Bause:2020xzj,Golz:2021imq,Golz:2022alh}.

\subsection{Inclusive semileptonic and radiative decays\label{sec:inclusivetheory}}
The standard approach for calculating the inclusive $\Bbar\to X_c\ell^-\nub$ decay rate is the heavy-quark expansion (HQE). Using the optical theorem, the inclusive rate is expressed in terms of a forward matrix element of a product of two weak currents at different spacetime points. This is then treated with the heavy-quark/operator-product expansion that yields a series of local operators, organized by powers of $1/m_b$ \cite{Gambino:2020jvv}, where the leading term corresponds to the ``partonic rate'' and terms up to order $1/m_b^5$ are known at tree level. The hadronic matrix elements are fitted to experimental data~\cite{Bordone:2021oof,Bernlochner:2022ucr} (in principle, they can also be calculated using lattice QCD \cite{Kronfeld:2000gk,Gambino:2017vkx,FermilabLattice:2018est}), and the matching coefficients are calculated perturbatively. For the partonic rate, the full kinematic distribution is known at order $\alpha_s^2$, and the total rate at $\alpha_s^3$ \cite{Fael:2020tow}. The full kinematic distribution of the $1/m_b^2$ contribution and the total rate at order $1/m_b^3$ have been calculated at order $\alpha_s$ \cite{Alberti:2012dn,Alberti:2013kxa,Mannel:2021zzr}. The theory uncertainty in inclusive $|V_{cb}|$ determinations has reached the 1\% level \cite{Bordone:2021oof}. Looking ahead, the proliferation of HQE parameters at higher orders can be reduced by considering reparametrization-invariant observables, in particular $q^2$ moments \cite{Fael:2018vsp}, as was already implemented in Ref.~\cite{Bernlochner:2022ucr}. Higher-order $\alpha_s$ corrections can be calculated, and improvements in the heavy-quark mass determination as well as novel heavy-quark mass schemes may be beneficial. At sub-percent level, possible duality violations and problems of HQE convergence \cite{MannelSnowmass} as well as QED corrections should be investigated further. The $b\to u\ell^-\nub$ and $b\to c\tau^-(\to\ell^-\nu\nub)\nub$ backgrounds also need to be treated carefully \cite{Mannel:2021mwe}.

The determination of $|V_{ub}|$ from inclusive $\Bbar \to X_u \ell^-\nub$ measurements is substantially more complicated due to the large charm background. Cutting away the $b\to c\ell^-\nub$ contribution with a requirement on the lepton energy leaves only the endpoint region with $2 E_\ell/m_b\sim 1$, where the local HQE breaks down. In this region, one needs to use a light-cone OPE, such that the HQE parameters are replaced by nonlocal matrix elements, the so-called shape functions \cite{Bauer:2001mh,Bauer:2002yu}. The shape functions can be evaluated through a combination of fits to the differential data and QCD-based modeling; see Ref.~\cite{Gambino:2020jvv} for details and prospects for improvements.

Also very important are the inclusive rare decays $\Bbar\to X_s\gamma$ and $\Bbar\to X_s\ell^+\ell^-$ \cite{Rare-decay-Overview-Whitepaper}, which can provide tight constraints on NP with different systematic uncertainties compared to the exclusive rare $b$ decays. The inclusive $\Bbar\to X_s\gamma$ branching fraction currently has a 5\% theory uncertainty \cite{Misiak:2020vlo}, while Belle~II may reach 2\% as shown in Tab.~\ref{tab:sensitivity}. The theory uncertainty can be reduced further by completing the NNLO QCD corrections without interpolation in the charm mass and by controlling nonperturbative effects that are expected to give few-\% contributions. For $\Bbar\to X_s\ell^+\ell^-$ decays at low $q^2$ the situation is similar \cite{Huber:2020vup}. At high $q^2$ the uncertainties are larger due to the breakdown of the HQE, as with $\Bbar\to X_u \ell^-\nub$ decays. At high $q^2$ it is advantageous to normalize the $\Bbar\to X_s\ell^+\ell^-$ rate to the $\Bbar \to X_u \ell^-\nub$ rate with the same kinematic cut \cite{Ligeti:2007sn}.

Finally, significant progress in lattice QCD has been made in direct computations of the forward matrix elements of the two weak currents that are needed to describe inclusive decay rates. Extracting the hadronic tensor from a Euclidean four-point function requires solving an inverse Laplace transform. This is an ill-posed problem in principle, but the severity of the problem can be reduced by limiting the energy resolution, which at the same time controls finite-volume effects \cite{Hansen:2017mnd,Hansen:2019idp,Gambino:2020crt,DeGrand:2022lmc,JaySnowmass} (see also Ref.~\cite{Hashimoto:2017wqo} for earlier work). Exploratory computations of inclusive semileptonic decay rates, along with a comparison to OPE predictions, have been performed successfully at a lower-than physical $b$-quark mass \cite{Gambino:2020crt,Maechler:2021kax,Gambino:2022dvu}. This paves the way for further calculations with physical parameters and controlled systematic uncertainties.

\subsection{Neutral-meson mixing}
The mass differences $\Delta m_d$ and $\Delta m_s$ in \Bz-\Bzb and \Bs-\Bsb mixing are known from experimental measurements with an extraordinary precision of $0.4\%$ and $0.03\%$, respectively \cite{HFLAV}. The theoretical uncertainties are at the $10\%$ level~\cite{Dowdall:2019bea,DiLuzio:2019jyq}. Reducing them will therefore have a significant impact on CKM constraints and on many NP models, including models proposed as explanations of the $b\to s\ell^+\ell^-$ anomalies.

The theoretical description of $\Delta m_d$ and $\Delta m_s$ at leading-order in weak effective theory involves the hadronic matrix elements $\langle \Bzb_{(s)} | Q_i | \Bz_{(s)} \rangle$ of five local four-fermion $\Delta B=2$ operators $Q_i$, of which only $Q_1$ contributes in the SM (see the caption of Fig.~\ref{fig:NPscales}). These matrix elements can be calculated using lattice QCD \cite{Dalgic:2006gp,Gamiz:2009ku,ETM:2013jap,Aoki:2014nga,FermilabLattice:2016ipl,Dowdall:2019bea} or sum rules \cite{Grozin:2016uqy,Grozin:2017uto,Kirk:2017juj,Grozin:2018wtg,King:2019lal}. Lattice results for the $SU(3)$-breaking ratio $\xi$, formed using the ratio of \Bs- and \Bz-meson mixing parameters, have reached percent-level precision~\cite{Gamiz:2009ku,Albertus:2010nm,Bazavov:2012zs,ETM:2013jap,Aoki:2014nga,FermilabLattice:2016ipl,Boyle:2018knm,Dowdall:2019bea}. For the mixing parameters themselves, there are currently some tensions among the lattice results from different groups employing different renormalization schemes, lattice discretizations, and numbers of dynamical quark flavors that need to be understood and resolved. It is likely possible to achieve sub-percent uncertainties within the next five years, at which point QED effects become important \cite{Boyle:2022uba}. The sum rules estimates can also be further improved by determining $1/m_b$ corrections to the strict HQET limit, by determining NNLO-QCD corrections to the QCD-HQET matching or by considering the sum rule in full QCD \cite{CKM-Overview-Whitepaper}.

For the width difference $\Delta\Gamma_s$, a large contribution to the theory uncertainty is due to the hadronic matrix elements of dimension-7 operators. A first lattice-QCD calculation of these operators has been reported in Ref.~\cite{Davies:2019gnp}, leading to a SM prediction of $\Delta\Gamma_s=0.092\pm0.014\ps^{-1}$, which can be compared to the current experimental average of $0.084 \pm 0.005\ps^{-1}$~\cite{HFLAV}. Hence, further substantial improvements in the theory uncertainties would have a big impact, particularly given that the experimental uncertainties at the LHC are projected to decrease even further~\cite{LHCb:2018roe,AtlasCMS-Whitepaper}.

In \Dz-\Dzb mixing, the \CP-violating contributions can be described reasonably well by local hadronic matrix elements of $\Delta C=2$ operators. These matrix elements have already been determined via sum rules~\cite{Kirk:2017juj} and lattice QCD~\cite{Carrasco:2014uya,Carrasco:2015pra,Bazavov:2017weg} with uncertainties of order 5-10\%, and further improvements are possible using standard methods in the coming years. However, the overall mixing process is dominated by long-distance effects, corresponding to hadronic matrix elements of two $\Delta C=1$ operators at different spacetime points. Analogous nonlocal matrix elements have been computed on the lattice for the kinematically simpler case of kaon mixing \cite{Christ:2012se,Bai:2014cva,Christ:2015pwa}, but substantial further theoretical and algorithmic developments are needed to extend this work to the case of charm mixing. It may be possible to employ novel methods similar to those developed for lattice calculations of inclusive decay rates (\cf Sec.~\ref{sec:inclusivetheory}) \cite{USQCD:2019hyg}.

\subsection{Heavy-hadron lifetimes}
Comparisons between theory and experiment for heavy-hadron lifetimes, or lifetime ratios, can also provide useful constraints on physics beyond the SM. Examples are the constraints on beyond-SM explanations of the $R(\DorDs)$ anomalies from $\tau_{B^+_c}$ \cite{Alonso:2016oyd}, and constraints from $\tau_{\Bs}/\tau_{\Bz}$ on NP in $b\to s\tau^+\tau^-$, on $B$-meson-based baryogenesis, and on NP in nonleptonic decays such as $b\to c\bar{c}s$ \cite{Lenz:2021bkv}. A recent experimental surprise is the LHCb measurement of the \Omegac lifetime, which rearranged the hierarchy of the charm-baryon lifetimes \cite{LHCb:2018nfa,LHCb:2019ldj,LHCb:2021vll,Gratrex:2022xpm,Belle-II:2022plj}.

Similar to inclusive semileptonic decays, the standard theoretical tool is the HQE (and some of the Wilson coefficients and hadronic matrix elements are shared between these applications). Significant sources of theoretical uncertainty for $b$-hadron lifetimes are spectator effects described by four-quark operators, and the Darwin term~\cite{Lenz:2020oce,Mannel:2020fts,Lenz:2021bkv,King:2021jsq,Gratrex:2022xpm,Lenz:2022rbq}. There have been exploratory lattice calculations of spectator effects more than 20 years ago
\cite{Becirevic:2001fy,Becirevic:2000sj,Flynn:2000hx,DiPierro:1999tb,DiPierro:1998ty}. New state-of-the-art lattice calculations would be desirable to complement sum-rule calculations.

\subsection{Nonleptonic bottom-hadron decays}
Nonleptonic bottom-hadron decays are widely used to study \CP violation, both in the SM and beyond. As discussed in Sec.~\ref{sec:CKMtests}, the CKM angles $\alpha$, $\beta$, and $\gamma$ can be determined from combinations of multiple nonleptonic $B_{(s)}$ decay modes that allow the dominant hadronic matrix elements to be obtained from the experimental data. However, except for the case of $\gamma$ \cite{Brod:2013sga}, subleading effects will no longer be negligible at the increased precision expected from experiments in the coming years, and need to be understood theoretically. An example is the $V_{ub}$ (``penguin'') contamination in $\Bz\to \jpsi\KS$ and $\Bs\to\jpsi\phi$, which affects determination of the CKM angles $\beta$ and $\beta_s$.

Flavor-symmetry relations are widely used in theoretical studies of nonleptonic $b$ decays \cite{Zeppenfeld:1980ex,Gronau:1990ka,Gronau:1994rj,Gronau:1995hm,Ciuchini:2011kd}. Calculating hadronic matrix elements for nonleptonic $b$ decays directly is very challenging. Commonly used approaches are perturbative QCD, QCD factorization/soft-collinear effective theory, and light-cone sum rules, often combined with an expansion in $1/m_b$ \cite{Beneke:1999br,Beneke:2000ry,Keum:2000wi,Keum:2000ph,Lu:2000em,Beneke:2001ev,Bauer:2001yt,Beneke:2003zv,Bauer:2004tj,Beneke:2004bn,Feldmann:2004mg,Ali:2007ff,Frings:2015eva,Huber:2016xod,Zhou:2016jkv,Wang:2017hxe,Huber:2020pqb,Bell:2020qus,Hua:2020usv,Huber:2022vix,Lu:2022fgz}. Recently, factorization was also extended to include QED \cite{Beneke:2020vnb,Beneke:2021jhp}.
Naturally, one may ask whether lattice-QCD calculations of nonleptonic $b$-hadron decays are possible. The lattice methods developed for exclusive decays with two or three hadrons in the final state \cite{Luscher:1986pf,Luscher:1991cf,Lellouch:2000pv,Lin:2001ek,Christ:2005gi,Hansen:2012tf,Briceno:2014uqa,Briceno:2015csa,Agadjanov:2016fbd,Briceno:2021xlc,Muller:2020wjo,Hansen:2021ofl} are insufficient for processes like $B\to \pi\pi$ due to the high center-of-mass energy of the mesons in the final state, at which rescattering to many different states with more than three hadrons would affect the finite-volume matrix elements. However, lattice calculations can contribute in other ways to the theory of nonleptonic $b$ decays, for example by providing first-principles predictions of the $B$-meson light-cone distribution amplitude \cite{Kawamura:2018gqz,Wang:2019msf,Zhao:2020bsx,Xu:2022krn}.

An interesting puzzle has emerged concerning the branching fractions of $B^0_{(s)} \to D^{\scalebox{0.4}{(}*\scalebox{0.4}{)-}}_{(s)}\{\pi^+,\,K^+\}$ decays, where improved QCD-factorization predictions are several standard deviations higher than experimental measurements \cite{Huber:2016xod,Bordone:2020gao}. The estimated theoretical uncertainties include, for the first time, the ${\cal O}(\Lambda_{\rm QCD}/m_b)$ corrections. The decays considered include some of the simplest, color-allowed processes that are unaffected by penguin or annihilation contributions, and it appears unlikely that large nonfactorizable effects can explain the deviations \cite{GubernariNonleptonicB}. Quasi-elastic rescattering effects were studied in Ref.~\cite{Endo:2021ifc}, where it was found that such effects cannot simultaneously account for the measured branching ratios of the color-allowed and color-suppressed decays. Beyond-the-SM explanations of the puzzle have been explored, for example, in Refs.~\cite{Iguro:2020ndk,Cai:2021mlt}, and may lead to large enhancements to \CP asymmetries that could be observed experimentally~\cite{Gershon:2021pnc}. Significant deviations from the SM predictions using QCD factorization were recently also found in dedicated measurements of the ratios between $\Gamma(\Bzb\to D^{*+}h^-)$ and $\mathrm{d}\Gamma(\Bzb\to D^{*+}\ell^-\nub)/\mathrm{d}q^2|_{q^2=m_h^2}$ \cite{Belle:2022afp}. 

While much of the work to date has focused on processes with $B_{(s)}$ mesons, the increased statistics from the continued running of the LHC will likely allow to observe \CP violation also in $b$-baryon decays. A non-vanishing \CP asymmetry is a measure of direct \CP violation, as baryons and antibaryons do not undergo mixing because of baryon number conservation. Theoretical studies of \CP violation in $b$-baryon decays can be found, \eg, in Refs.~\cite{Lu:2009cm,Hsiao:2014mua,Gronau:2013mza,He:2015fwa,Roy:2019cky,Dery:2020lbc,Roy:2020nyx,Bensalem:2002pz,Durieux:2016nqr}.

\subsection{Nonleptonic charm-hadron decays}
In charm decays, the size of diagrams from penguin operators is determined by the ratio $m_b/m_W\ll 1$, which is much smaller than in the case of bottom decays, where the relevant ratio is $m_t/m_W>1$. It follows that penguin operators are usually not relevant for charm decays, and the GIM mechanism is extremely effective. In singly Cabibbo-suppressed decays, \CP violation in the SM is proportional to the small non-unitary contribution of the relevant $2\times 2$ submatrix of the CKM matrix.

Direct charm \CP violation has been observed by LHCb in a difference of direct \CP asymmetries, $\Delta a_{\CP}^{\rm dir} \equiv a_{\CP}^{\rm dir}(D^0\rightarrow K^+K^-) - a_{\CP}^{\rm dir}(D^0\rightarrow \pi^+\pi^-)$, with the result $(-0.161 \pm 0.028)\%$ \cite{HFLAV}. The SM prediction is of order $\Delta a_{\CP}^{\rm dir}\sim 10^{-3} \times r_{\mathrm{QCD}}$, with $r_{\mathrm{QCD}}$ being a ratio of pure low-energy QCD amplitudes. Calculating the QCD amplitudes from first principles is even more challenging than in the case of nonleptonic $b$ decays, due to the stronger QCD coupling at the lower energy and the lower heavy-quark mass, meaning that many of the theoretical tools discussed in the previous Section are less suitable. Depending on the methods used to estimate $r_{\mathrm{QCD}}$, one arrives at very different interpretations of the result \cite{Grossman:2019xcj, Brod:2011re, Soni:2019xko, Schacht:2021jaz, Khodjamirian:2017zdu, Chala:2019fdb, Li:2019hho, Bediaga:2022sxw}. Further work is clearly needed. The theoretical understanding can also be improved by combining measurements of the \CP asymmetries in all singly-Cabibbo-suppressed charm-meson decays, taking advantage of flavor-$SU(3)$ sum rules.  In addition, the long-term prospects for direct lattice-QCD calculations of the relevant QCD amplitudes using the Lellouch-L\"uscher approach \cite{Luscher:1986pf,Luscher:1991cf,Lellouch:2000pv,Lin:2001ek,Christ:2005gi,Hansen:2012tf,Briceno:2014uqa,Briceno:2015csa,Agadjanov:2016fbd,Briceno:2021xlc,Muller:2020wjo,Hansen:2021ofl} are better than for nonleptonic $B$ decays, due to the smaller number of open multi-hadron channels at $\sqrt{s}\sim m_D$.

Like in the bottom sector, there are also interesting opportunities to study charm \CP violation in decays of baryons. Theoretical aspects are discussed in Refs.~\cite{Kang:2010td,Bigi:2012ev,Cheng:2018hwl,Grossman:2018ptn,Wang:2019dls,Shi:2019vus}.

\subsection{Model building\label{sec:BSM}}
The ultimate goal of the efforts discussed in this report is to find clear manifestations of physics beyond the SM, and to narrow down the structure and parameters of what theory will replace the SM. The efforts to construct new candidate theories, \ie, beyond-SM model building, are discussed from a general point of view in the Theory-Frontier topical-group-8 report \cite{TF08-Report} and, in the context of flavor physics, in Refs.~ \cite{Altmannshofer:2022aml,LFUV-Overview-Whitepaper}.

Beyond-SM model building may be approached from different directions. Many models are primarily designed to address questions relating to naturalness problems, dark matter, or baryogenesis. Nevertheless, such models often lead to new sources of flavor-changing interactions that may be observed in weak decays of $b$ or $c$ quarks, leading to tight constraints. On the other hand, the observation of deviations from the SM in weak decays of $b$ or $c$ quarks motivates a directed effort to build models that can explain these deviations while remaining consistent with other measurements. Typically, the measurements are initially interpreted in a model-independent way at the level of Wilson coefficients in a low-energy effective theory. If the scale of NP is far above the weak scale, the low-energy effective NP operators must be invariant under the SM gauge group. This condition imposes interesting new relations among different processes \cite{Bhattacharya:2014wla,Alonso:2014csa,Barbieri:2015yvd,Buttazzo:2017ixm,Bordone:2017anc}. The next level could be simplified models with new particles capable of producing the observed deviations, but not subject to the requirement of providing a renormalizable, self-consistent theory. Finding an ultraviolet completion would then be the next step.

Many of the models constructed to explain the $b\to s\ell^+\ell^-$ anomalies contain $Z^\prime$ bosons \cite{Altmannshofer:2014cfa,Crivellin:2015lwa,Celis:2015ara,Fuyuto:2015gmk, Bonilla:2017lsq,Bhatia:2017tgo,Alonso:2017uky,King:2017anf,Allanach:2018lvl,Becirevic:2018afm,Altmannshofer:2019xda,Bhatia:2021eco,Greljo:2021xmg,Greljo:2021npi,Niehoff:2015bfa,Sannino:2017utc,Carmona:2017fsn,Chung:2021ekz,Marzocca:2021azj,Becirevic:2022tsj}, typically with masses of a few \tev, and therefore within reach of the LHC or future colliders \cite{Greljo:2017vvb,Allanach:2017bta,Abdullah:2017oqj,Kohda:2018xbc, Allanach:2019mfl,Huang:2021nkl}. Another possibility are leptoquarks \cite{Hiller:2014yaa,Gripaios:2014tna,Alonso:2015sja, Bauer:2015knc,Calibbi:2015kma,Fajfer:2015ycq,Barbieri:2015yvd,Barbieri:2016las,Bhattacharya:2016mcc,Hiller:2016kry,Becirevic:2016oho,Cai:2017wry,Crivellin:2017zlb,Assad:2017iib,Buttazzo:2017ixm,Angelescu:2018tyl,Marzocca:2018wcf,Popov:2019tyc,Cornella:2019hct,Fuentes-Martin:2020hvc,FileviezPerez:2022rbk}, which can arise in many different beyond-SM scenarios, including in composite models \cite{Gripaios:2014tna,Barbieri:2016las,Marzocca:2018wcf}, in the MSSM with $R$-parity violation (in the form of the sbottom squark) \cite{Deshpande:2016yrv, Das:2017kfo, Altmannshofer:2017poe, Earl:2018snx, Trifinopoulos:2018rna}, and as gauge bosons models in models with enlarged gauge groups \cite{DiLuzio:2017vat,Bordone:2017bld,Barbieri:2017tuq,Calibbi:2017qbu, Greljo:2018tuh,Blanke:2018sro,Balaji:2018zna,Fornal:2018dqn,Heeck:2018ntp,FileviezPerez:2013zmv}. Several models can simultaneously explain the $b\to s\ell^+\ell^-$ and $b\to c\tau^- \nub$ anomalies, and/or simultaneously address other questions, for example by providing a dark-matter candidate. Interestingly, if requiring that a single mediator is responsible for the effects in both $b\to s\ell^+\ell^-$ and $b\to c\tau^-\nub$ at tree level, there is a unique choice of leptoquark: the $SU(3)$-triplet, $SU(2)$-singlet, hypercharge-$2/3$ vector leptoquark, usually denoted as $U_1$. This leptoquark may be one of the gauge bosons in a generalized version \cite{DiLuzio:2017vat,Bordone:2017bld,Bordone:2018nbg,Fuentes-Martin:2022xnb,Barbieri:2019zdz} of the Pati-Salam grand unified theory \cite{Pati:1974yy}.

\section{Current and future U.S.\ involvement\label{sec:USinvolvement}}
The U.S.\ has been a leader in heavy-quark physics, involving a vigorous community and a series of extremely successful domestic experiments. The CLEO experiment at Cornell and the BaBar experiment at SLAC have been fundamental in today's understanding of $B$ and charm physics, and the CDF and D0 experiments at the Tevatron have pioneered the study of \Bs mesons and $b$-baryons in addition to set the bases for precise heavy-quark physics at hadron colliders. Such a strong domestic program did not limit participation in offshore experiments, such as Belle at KEK in Japan. Since the shutdown of PEP-II and of the Tevatron, the U.S. heavy-flavor community has exclusively relied on offshore experiments, namely LHCb, Belle II and BESIII. About 60 physicists from six U.S. institutions participate to the LHCb collaboration. U.S. groups have lead the design, construction and commissioning of the upstream silicon tracker for Upgrade I; the development of trigger algorithms for real-time analysis; and several key physics measurements, including searches for lepton-flavor-universality violation, \CP-violation in the \Bs system, and discovery of tetra- and penta-quark states. The Belle II collaboration includes about 80 physicists from 17 U.S. institutions. BNL hosts a Tier 1 GRID computing facility for Belle II. U.S. groups have played major roles in the design, construction, commissioning, and operations of the Belle II hadron- and muon-identification detectors; in critical machine-detector interface studies of beam backgrounds; in management and leadership positions, and in data analysis. The BESIII collaboration currently includes six physicists from three US institutions. The BESIII U.S. groups contribute to data analysis and have senior members holding significant leadership positions, such as in the publication committee. The combined physics output of LHCb, Belle II and BESIII is in the order of 220 publications per year\footnote{With Belle II accumulating more and more data this number is expected to increase by at least 20\% over the next decade.} -- comparable to experiments at the energy frontier, which typically rely on much larger funding and U.S. participation.

International recognition of the importance of a continued heavy-flavor-physics program in the next decades is evident from the commitments in Europe and Asia. An example is the strong support given by the 2020 European Strategy Update to the study of heavy-flavor physics at the HL-LHC. For the U.S. HEP program to have the breadth to assure meaningful role in future discoveries, support for a significant U.S. participation in future heavy-flavor experiments needs to be assured. U.S. groups can have leading roles in the design and construction of detector and data-acquisition-system upgrades planned over the next ten years. Hence, U.S. contributions to LHCb Upgrade II and future upgrades of Belle II must be encouraged.

The successful experience of the LHC has demonstrated that experiments at energy-frontier facilities can be effectively complemented by experiments focusing on indirect searches for beyond-SM physics through the study of weak decays of $b$ and $c$ quarks. While the identification of the next energy-frontier facility will be mostly motivated by the need to understand the mechanism behind the electroweak-symmetry breaking and/or by the need to increase the reach of direct searches, the important role of heavy-quark physics should still be considered as key for a broad and rich HEP program. Strong U.S. participation in these efforts should also be encouraged.

The experimental progress in heavy-flavor physics has often benefited from a close collaboration with the theory community. The U.S. has strong theory groups working on quark-flavor physics, which are internationally recognized and influential. Scientists in the U.S. have pioneered heavy-quark effective theory, nonrelativistic QCD, heavy-hadron chiral perturbation theory, and lattice formulations for heavy quarks. Lattice gauge theory plays a crucial role for the physics program discussed in this report, and the U.S. has a very active lattice community. In flavor physics, the Fermilab Lattice, HPQCD, MILC, and RBC-UKQCD Collaborations are particularly influential. The majority of lattice-field theory researchers in the U.S. are also members of the USQCD collaboration, which was founded in 1999 for the purpose of creating and utilizing software and dedicated hardware resources for lattice gauge theory calculations. As of April 2022, USQCD has 169 members. USQCD software is open-source and used widely by the worldwide community. In order for the strong theory and lattice efforts in the U.S. to continue, stable support for researchers and for computing resources is essential.

In the next decades, weak decays of $b$ and $c$ quarks will offer a unique opportunity to reveal NP that is not directly accessible at the energy frontier. A healthy U.S. HEP program will endeavor to be among the leaders in this research.

\section*{Acknowledgments\addcontentsline{toc}{section}{Acknowledgments}}
We gratefully acknowledge the authors of the contributed white papers and all individuals who attended the RF1 workshops. We thank Flavio Archilli, Wolfgang Altmannshofer, Bhubanjyoti Bhattacharya, Thomas E.\ Browder, Alakabha Datta, Paolo Gambino, Xian-Wei Kang, Lingfeng Li, James~F.~Libby, Alexander Lenz, Dominik~S.~Mitzel, Dean Robinson, Mike Roney, Matthew~S.\ Rudolph, Alan~J.~Schwartz, Phillip Urquijo, Oliver Witzel, and Roman Zwicky for helpful comments on this report.


\bibliographystyle{JHEP}
\bibliography{Rare/RF01/references}
\addcontentsline{toc}{section}{References}





%% file: Rare/RF01/macros.tex
\newcommand{\todo}[1]{\textcolor{red}{#1}}
\newenvironment{todoenv}{\par\color{red}}{\par}

\def\CP{{\ensuremath{C\!P}}\xspace}
\newcommand{\ACP}{\ensuremath{A_{\CP}}\xspace}
\newcommand{\SCP}{\ensuremath{S_{\CP}}\xspace}
\newcommand{\BF}{\ensuremath{\mathcal{B}}\xspace}

\newcommand{\nospaceunit}[1]{\ensuremath{{\rm #1}}}
\newcommand{\aunit}[1]{\ensuremath{{\rm\,#1}}}
\newcommand{\unit}[1]{\aunit{#1}\xspace}

\def\degrees{\ensuremath{^{\circ}}\xspace}
\def\rad{\aunit{rad}}
\def\mrad{\aunit{mrad}}

\newcommand{\tev}{\aunit{Te\kern -0.1em V}\xspace}
\newcommand{\gev}{\aunit{Ge\kern -0.1em V}\xspace}
\newcommand{\mev}{\aunit{Me\kern -0.1em V}\xspace}
\newcommand{\kev}{\aunit{ke\kern -0.1em V}\xspace}
\newcommand{\ev}{\aunit{e\kern -0.1em V}\xspace}
\newcommand{\mevc}{\ensuremath{\aunit{Me\kern -0.1em V\!/}c}\xspace}
\newcommand{\gevc}{\ensuremath{\aunit{Ge\kern -0.1em V\!/}c}\xspace}
\newcommand{\mevcc}{\ensuremath{\aunit{Me\kern -0.1em V\!/}c^2}\xspace}
\newcommand{\gevcc}{\ensuremath{\aunit{Ge\kern -0.1em V\!/}c^2}\xspace}
\newcommand{\gevgevcc}{\ensuremath{\gev^2/c^2}\xspace} 
\newcommand{\gevgevcccc}{\ensuremath{\gev^2/c^4}\xspace} 

\def\km   {\aunit{km}\xspace}
\def\m    {\aunit{m}\xspace}
\def\ma   {\ensuremath{\aunit{m}^2}\xspace}
\def\cm   {\aunit{cm}\xspace}
\def\cma  {\ensuremath{\aunit{cm}^2}\xspace}
\def\mm   {\aunit{mm}\xspace}
\def\mma  {\ensuremath{\aunit{mm}^2}\xspace}
\def\mum  {\ensuremath{\,\upmu\nospaceunit{m}}\xspace}
\def\muma {\ensuremath{\,\upmu\nospaceunit{m}^2}\xspace}
\def\nm   {\aunit{nm}\xspace}
\def\fm   {\aunit{fm}\xspace}
\def\barn {\aunit{b}\xspace}
\def\mbarn{\aunit{mb}\xspace}
\def\mub  {\ensuremath{\,\upmu\nospaceunit{b}}\xspace}
\def\nb   {\aunit{nb}\xspace}
\def\invnb{\ensuremath{\nb^{-1}}\xspace}
\def\pb   {\aunit{pb}\xspace}
\def\invpb{\ensuremath{\pb^{-1}}\xspace}
\def\fb   {\ensuremath{\aunit{fb}}\xspace}
\def\invfb{\ensuremath{\fb^{-1}}\xspace}
\def\ab   {\ensuremath{\aunit{ab}}\xspace}
\def\invab{\ensuremath{\ab^{-1}}\xspace}

\def\sec  {\ensuremath{\aunit{s}}\xspace}
\def\ms   {\ensuremath{\aunit{ms}}\xspace}
\def\mus  {\ensuremath{\,\upmu\nospaceunit{s}}\xspace}
\def\ns   {\ensuremath{\aunit{ns}}\xspace}
\def\ps   {\ensuremath{\aunit{ps}}\xspace}
\def\fs   {\aunit{fs}}
\def\mhz  {\ensuremath{\aunit{MHz}}\xspace}
\def\khz  {\ensuremath{\aunit{kHz}}\xspace}
\def\hz   {\ensuremath{\aunit{Hz}}\xspace}
\def\invps{\ensuremath{\ps^{-1}}\xspace}
\def\invns{\ensuremath{\ns^{-1}}\xspace}

\makeatletter
\ifcase \@ptsize \relax
  \newcommand{\miniscule}{\@setfontsize\miniscule{4}{5}}
\or
  \newcommand{\miniscule}{\@setfontsize\miniscule{5}{6}}
\or
  \newcommand{\miniscule}{\@setfontsize\miniscule{5}{6}}
\fi
\makeatother
\DeclareRobustCommand{\optbar}[1]{\shortstack{{\miniscule (\rule[.5ex]{1.25em}{.18mm})}
  \\ [-.7ex] $#1$}}

\def\Fbar {{\kern 0.2em\overline{\kern -0.2em F}{}}\xspace}

\def\Kbar {{\kern 0.2em\overline{\kern -0.2em K}{}}\xspace}
\def\KS   {\ensuremath{K^0_{\scriptscriptstyle\rm S}}\xspace} 
\def\KL   {\ensuremath{K^0_{\scriptscriptstyle\rm L}}\xspace}

\def\Dbar {\kern 0.2em\overline{\kern -0.2em D}{}\xspace}
\def\Dz   {\ensuremath{D^0}\xspace}
\def\Dzb  {\ensuremath{\Dbar^0}\xspace}
\def\Ds   {{\ensuremath{D_s}}\xspace}
\def\Dsp  {{\ensuremath{D^+_s}}\xspace}
\def\Dsm  {{\ensuremath{D^-_s}}\xspace}
\def\Dss  {{\ensuremath{D^{*}_s}}\xspace}
\def\Dssp {{\ensuremath{D^{*+}_s}}\xspace}
\def\Dssm {{\ensuremath{D^{*-}_s}}\xspace}
\def\DorDsm{{\ensuremath{D^{\scalebox{0.4}{(}*\scalebox{0.4}{)}-}}}\xspace}
\def\DsorDssm{{\ensuremath{D^{\scalebox{0.4}{(}*\scalebox{0.4}{)}-}_{s}}}\xspace}
\def\DorDs{{\ensuremath{D^{\scalebox{0.4}{(}*\scalebox{0.4}{)}}}}\xspace}
\def\DsorDss{{\ensuremath{D^{\scalebox{0.4}{(}*\scalebox{0.4}{)}}_{s}}}\xspace}

\def\Bbar {\kern 0.18em\overline{\kern -0.18em B}{}\xspace}
\def\Bz   {\ensuremath{B^0}\xspace}
\def\Bzb  {\ensuremath{\Bbar^0}\xspace}
\def\Bs   {\ensuremath{B_s^0}\xspace}
\def\Bsb  {\ensuremath{\Bbar_s^0}\xspace}

\mathchardef\PLambda="7103
\def\Lbar {\kern 0.2em\overline{\kern -0.2em \PLambda}{}\xspace}
\def\Lz   {{\ensuremath{\PLambda^0}}\xspace}
\def\Lc   {{\ensuremath{\PLambda^+_c}}\xspace}
\def\Lcb  {{\ensuremath{\Lbar{}^-_c}}\xspace}
\def\Lb   {{\ensuremath{\PLambda^0_b}}\xspace}
\def\Lbb  {{\ensuremath{\Lbar{}^0_b}}\xspace}

\mathchardef\PXi="7104
\def\Xic  {{\ensuremath{\PXi_c}}\xspace}
\def\Xicp {{\ensuremath{\PXi^+_c}}\xspace}
\def\Xicz {{\ensuremath{\PXi^0_c}}\xspace}

\mathchardef\POmega="710A
\def\Omegac{{\ensuremath{\POmega^0_c}}\xspace}

\def\jpsi {{\ensuremath{{J\mskip -3mu/\mskip -2mu\psi\mskip 2mu}}}\xspace}

\def\nub  {{\ensuremath{\overline{\nu}}}\xspace}

\renewcommand{\eg}{\mbox{\itshape e.g.}\xspace}
\renewcommand{\ie}{\mbox{\itshape i.e.}\xspace}
\renewcommand{\etal}{\mbox{\itshape et al.}\xspace}
\renewcommand{\etc}{\mbox{\itshape etc.}\xspace}
\newcommand{\cf}{\mbox{\itshape cf.}\xspace}
\newcommand{\ffp}{\mbox{\itshape ff.}\xspace}
\newcommand{\vs}{\mbox{\itshape vs.}\xspace}